\documentclass[a4paper,11pt]{article}
 
 \usepackage[english]{babel}  
 \usepackage[utf8]{inputenc}

 \usepackage[totalwidth=16.0cm,totalheight=24cm]{geometry}
 \usepackage[dvipsnames]{xcolor}  
  
 \usepackage{multirow}
 
\usepackage{amsmath,amssymb,amsthm,mathtools,mathrsfs}
\usepackage{bm,faktor}

  \newcommand{\ga}{\alpha}

  \newcommand{\Cd}{\dot{C}}

  \newcommand{\gd}{\delta}

  \newcommand{\eh}{\hat{e}}					

  \renewcommand{\bf}{\bm{f}}				
  
  \newcommand{\fd}{\dot{f}}

  \newcommand{\Phih}{\hat{\Phi}}

  \newcommand{\cG}{\mathcal{G}}

  \newcommand{\fkgg}{\mathfrak{g}}

  \newcommand{\hd}{\dot{h}}

  \newcommand{\bh}{\bm{h}}

  \newcommand{\scrI}{\mathscr{I}}
  
  \newcommand{\lh}{\hat{l}}					

  \newcommand{\gl}{\lambda}
  
  \newcommand{\gL}{\Lambda}
  \newcommand{\gld}{\dot{\lambda}}

  \newcommand{\mb}{\bar{m}}

  \newcommand{\fkm}{\mathfrak{m}}

  \newcommand{\nh}{\hat{n}}

  \newcommand{\bn}{\bm{n}}

  \newcommand{\cO}{\mathcal{O}}				

  \newcommand{\go}{\omega}

  \newcommand{\cP}{\mathcal{P}}

  \newcommand{\bbR}{\mathbb{R}}
  \newcommand{\gr}{\rho}

  \newcommand{\gs}{\sigma}
  \newcommand{\gS}{\Sigma}

  \newcommand{\tb}{\bar{t}}					

  \newcommand{\cT}{\mathcal{T}}

  \newcommand{\gth}{\theta}					
  
  \newcommand{\gTh}{\Theta}

  \newcommand{\uh}{\hat{u}}					

    \newcommand{\gU}{\Upsilon}

  \newcommand{\zb}{\bar{z}}					

\newcommand{\from}{\colon}
\newcommand{\xto}[2][]{\xrightarrow[#1]{#2}}

\DeclareMathOperator{\ISO}{ISO\!}
\newcommand{\iso}{\mathfrak{iso}}
\DeclareMathOperator{\Carr}{Carr\!}
\newcommand{\carr}{\mathfrak{carr}}

\DeclareMathOperator{\SO}{SO\!}

\newcommand{\so}{\mathfrak{so}}

\newcommand{\Quotient}[2]{\faktor{#1}{#2}}

\newcommand{\covD}{\nabla}

\newcommand{\covDt}{\widetilde{\nabla}}
\newcommand{\parD}{\partial}

\newcommand{\LieD}{\mathcal{L}}
\newcommand{\intD}{\lrcorner}

\newcommand{\W}{\wedge}
\newcommand{\So}[1]{\Gamma\left[ #1 \right]}
\newcommand{\Co}[1]{\mathcal{C}^{\infty}\left( #1 \right)}

\newcommand{\Mtx}[1]{\begin{pmatrix} #1	\end{pmatrix}}
\newcommand{\sMtx}[1]{\begin{bmatrix} #1	\end{bmatrix}}  
  
\newcommand{\xeq}[1]{\stackrel{#1}{=}}

\usepackage{relsize}

\newtheorem{Proposition}{Proposition}[section]
\newtheorem{Theorem}{Theorem}[section]

\theoremstyle{definition}
\newtheorem{Definition}{Definition}[section]
\numberwithin{equation}{section}  

\usepackage{comment}
\usepackage{tcolorbox}
\excludecomment{Extra} 

\usepackage[backend=biber, isbn=false , url=false, hyperref=true, doi=true,giveninits=true,maxnames=4,style=numeric-comp,sorting=none,date=year]{biblatex}

\renewbibmacro{in:}{}
\DeclareFieldFormat[article]{volume}{\bibstring{volume}\addnbspace #1}
\DeclareFieldFormat[article]{number}{\bibstring{number}\addnbspace #1}
\AtEveryBibitem{\clearlist{language}\clearfield{month}}
\renewbibmacro*{journal+issuetitle}{%
	\usebibmacro{journal}%
	\setunit*{\addcomma\space}
	\iffieldundef{series}
	{}
	{\newunit
		\printfield{series}%
		\setunit{\addcomma\space}}
	\usebibmacro{volume+number+eid}%
	\setunit{\addspace}%
	\usebibmacro{issue+date}%
	\setunit{\addcolon\space}%
	\usebibmacro{issue}%
	\newunit}

\renewbibmacro*{volume+number+eid}{%
	\printfield{volume}%
	\setunit{\addcomma\space}
	\printfield{number}%
	\setunit{\addcomma\space}%
	\printfield{eid}}

\usepackage{hyperref}
\usepackage[all]{hypcap}
\begin{document} 
	\pagestyle{plain}
	\title{\textbf{Carrollian manifolds and null infinity:\\ A view from Cartan geometry}}
	\author{
		Yannick Herfray\footnote{Yannick.Herfray@umons.ac.be} \\
	\emph{Département Physique de l'Univers, Champs et Gravitation,UMONS, Belgique.}} 
	\date{}
	\maketitle
\begin{abstract}
We discuss three different (conformally) Carrollian geometries and their relation to null infinity from the unifying perspective of Cartan geometry. Null infinity \emph{per se} comes with numerous redundancies in its intrinsic geometry and the two other Carrollian geometries can be recovered by making successive choices of gauge. This clarifies the extent to which one can think of null infinity as being a (strongly) Carrollian geometry and we investigate the implications for the corresponding Cartan geometries. 

The perspective taken, which is that characteristic data for gravity at null infinity are equivalent to a Cartan geometry for the Poincaré group, gives a precise geometrical content to the fundamental fact that ``gravitational radiation is the obstruction to having the Poincaré group as asymptotic symmetries''.

\end{abstract}
\maketitle

\section{Introduction/Summary}

\begin{figure}\label{Galilean versus Carrollian manifold}
	\begin{center}
		\includegraphics[width= 0.7\linewidth]{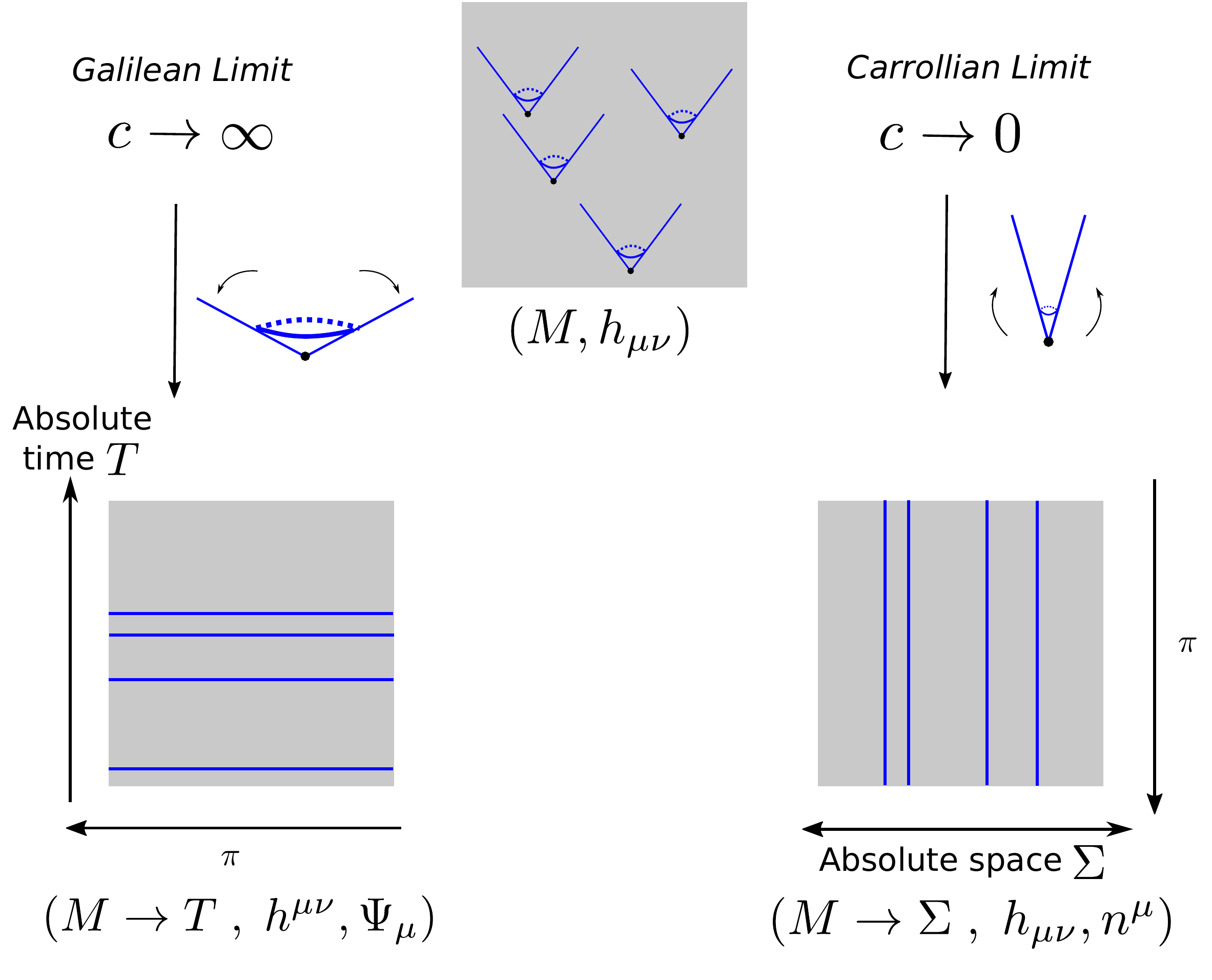}
	\end{center}
	\caption{Galilean (non-relativistic) versus Carrollian (ultra-relativistic) limit:
	As the speed of light goes to infinity/zero the tangent null cones degenerate into lines/hypersurfaces of constant absolute space/time }
\end{figure}

Carrollian geometries naturally emerge as the ultra-relativistic\footnote{If $v$ is the typical velocity of the system, one can easily argue that ``ultra-relativistic'' should be kept for situations where $c/v \to 1$ (resulting in conformal geometry where only the causal structure is preserved). From this perspective the Carrollian limit $c/v \to 0$ should perhaps be called the ultra-local limit. } limit $c \to 0$ of pseudo-Riemannian manifolds. \cite{levy-leblond_nouvelle_1965,bacry_possible_1968}. This parallels the appearance of Galilean geometry in the non-relativistic limit $c \to \infty$ and in some sense these are dual to each others \cite{duval_carroll_2014}. Loosely speaking, while the Galilean limit forces null cones to spread out into degenerate planes of constant absolute time, in the Carrollian limit null cones collapse into degenerate lines of constant absolute space (figure \ref{Galilean versus Carrollian manifold}).

Accordingly, the fundamental building block of any Carrollian geometry consists of a manifold $\scrI$ foliated by lines. The space of lines $\gS$ typically is a manifold and this results in a fibre bundle $\scrI \xto{\pi} \gS$. One obtains a Carrollian geometry $\left(\scrI \to \gS, n^{\mu}, h_{\mu\nu}\right)$ if the foliation $\scrI \to \gS$ is generated by a preferred vector field $n^{\mu}$ and is complemented by a degenerate metric $h_{\mu\nu}$ with one-dimensional kernel spanned by $n^{\mu}$.

This type of structure naturally appears in many different contexts of physical interest. They typically result of a limiting process or appear as the geometry induced on a null hypersurface. Applications include Hamiltonian analysis \cite{henneaux_geometry_1979}, Carrollian particles \cite{bergshoeff_dynamics_2014,marsot_planar_2021}, ultra-relativistic fluid mechanics \cite{de_boer_perfect_2018,ciambelli_flat_2018,ciambelli_covariant_2018,ciambelli_carrollian_2019,campoleoni_two-dimensional_2019,ciambelli_gauges_2020,de_boer_non-boost_2020}, cosmology \cite{de_boer_carroll_2021}, Carrollian  \cite{bergshoeff_carroll_2017,gomis_newton-hookecarrollian_2020,guerrieri_carroll_2021,perez_asymptotic_2021,henneaux_carroll_2021} and conformally Carrollian \cite{bagchi_field_2019,bagchi_field_2020,bagchi_bms_2021,banerjee_interacting_2021} field theories, higher-spin \cite{bergshoeff_three-dimensional_2017,ammon_scalar_2021,campoleoni_carrollian_2021}, super-symmetric \cite{bergshoeff_symmetries_2016,ravera_ads_2019,barducci_vector_2019,figueroa-ofarrill_kinematical_2019,ali_n-extended_2020,prabhu_novel_2021}  and non-commutative \cite{ballesteros_lorentzian_2020} extensions, null hypersurfaces and isolated horizons \cite{penrose_geometry_1972,ashtekar_isolated_2004,gourgoulhon_31_2006,duval_carroll_2017,hopfmuller_gravity_2017,hopfmuller_null_2018,chandrasekaran_symmetries_2018,donnay_carrollian_2019,penna_near-horizon_2019,oliveri_boundary_2020,chandrasekaran_anomalies_2021,chandrasekaran_brown-york_2021}, geometry of space/time-like infinities \cite{ashtekar_unified_1978,gibbons_ashtekar-hansen_2019, figueroa-ofarrill_carrollian_2021} and, finally, geometry of null infinity which is the main concern of this article.

It is a well-known fact that the conformal boundary of an asymptotically flat spacetime, which we will generically refer to as ``null infinity'', is a (conformal) Carrollian manifold. There is in fact a precise sense in which flat limits of an asymptotically AdS spacetime correspond to ultra-relativistic contractions of the boundary \cite{compere_lambda-bms_4_2019}.\\

 The gravitational characteristic data however induce on null infinity much more than a mere Carrollian geometry \cite{geroch_asymptotic_1977,ashtekar_radiative_1981,penrose_spinors_1984}. In four dimensions this geometry is especially rich and e.g. is closely related to asymptotic twistors \cite{penrose_twistor_1973,eastwood_edth-differential_1982} and Newman's H-space \cite{newman_heaven_1976,ko_theory_1981}.

 Depending on the perspective taken, especially the extent to which one wishes to preserve covariance and conformal invariance, there are various ways to describe the induced geometry. The purpose of this article is to discuss these in a coherent form emphasising their differences and relationships by framing them in the context of Cartan geometry.

Cartan geometry is a powerful organising tool in this context because the different views on the geometry of null infinity can in fact be captured by various homogeneous spaces which are all (conformal) Carrollian geometries with topology $\bbR \times S^2$. Cartan geometry amounts to making these models curved and thus capture the  corresponding local geometry (This is closely related to the ``gauging'' procedure common in physics, see \cite{wise_symmetric_2009,wise_macdowellmansouri_2010} for pedagogical physicist-oriented introductions and \cite{sharpe_differential_1997,cap_parabolic_2009} for standard references). These homogeneous models are simple enough to be presented in this introduction; the detailed discussion of the corresponding curved geometries will form the core of this article.
 
 Before discussing null infinity, it is instructive to consider the simpler model given by three-dimensional\footnote{In this introduction we restrict to 3 dimensions for concretenesses but in the core of the article we will treat uniformly all dimension $n\geq2$ as well as all the possible cosmological constant for the Carrollian geometries, see Table \ref{Table: introduction summary}.} homogeneous Carrollian--de Sitter spacetime \cite{bacry_possible_1968,figueroa-ofarrill_spatially_2019}:
\begin{equation}\label{Introduction: Carroll dS}
Carr_{dS}^{(3)} := \Quotient{\Carr_{dS}\left(3\right)}{\ISO\left(2\right)}.
\end{equation}
This is an homogeneous space for the Carroll--de Sitter group $\Carr_{dS}\left(3\right) := \bbR^3 \rtimes \SO\left(3\right)$ (with the stabiliser $\ISO\left(2\right)$ corresponding to a combination of rotations and Carrollian boosts). It can be obtained as the ultra-relativistic limit of $dS^{(3)}$, and therefore by construction is a Carrollian geometry. It is also equipped with a preferred compatible connection\footnote{i.e. satisfying $\covD h_{\mu\nu}=0$, $\covD n^{\mu}=0$.} $\covD$ and altogether this defines, in the terminology of \cite{duval_carroll_2014}, a strongly Carrollian geometry $\left(\scrI\to \gS , n^{\mu}, h_{\mu\nu}, \covD\right)$. In fact, as we will show in the first section of this article, strongly Carrollian geometries are in one-to-one correspondence with Cartan geometry modelled on \eqref{Introduction: Carroll dS}. What is more, we will obtain a result which applies uniformly to all dimensions $n\geq2$ and all possible cosmological constant i.e. Carroll--De Sitter, Carroll and Carroll--Anti De Sitter. Taking the cosmological constant to zero one recovers the results from \cite{hartong_gauging_2015}, restricting to dimension $2+1$ those of \cite{bergshoeff_three-dimensional_2017, ravera_ads_2019,matulich_limits_2019} and imposing a flatness condition the ones from \cite{figueroa-ofarrill_geometry_2019}.

 The model for null infinity is, without surprise, an homogeneous space for the (connected component of the) Poincaré group $\ISO\left(3,1\right) = \bbR^4 \rtimes \SO\left(3,1\right)_{0}$,
\begin{equation}\label{Introduction: Null infinity}
\scrI^{(3)} := \Quotient{\bbR^4 \rtimes \SO\left(3,1\right)_0}{\bbR^{3} \rtimes\left( \bbR^* \times \ISO\left(2\right)\right)}.
\end{equation}
The stabiliser now is a combination of rotations, Carrollian boosts and dilatations, corresponding to $\bbR^* \times \ISO\left(2\right)$, in semi-direct product with Carrollian special conformal transformations $\bbR^3$. There is an essential difference with the previous model: The appearance of $\SO\left(3,1\right)_0$, the conformal group of the two-sphere, means that this second model is only equipped with a \emph{conformal} Carrollian geometry \cite{duval_conformal_2014-1}. This well-known fact is related to the appearance of the BMS group \cite{duval_conformal_2014, ashtekar_geometry_2015, ciambelli_carroll_2019,prabhu_twistorial_2021} as symmetries of asymptotically flat spacetimes. On top of being conformally Carrollian, the model for null infinity \eqref{Introduction: Null infinity} is equipped with a Poincaré operator \cite{herfray_asymptotic_2020}: this is a differential operator $\cP$ acting on densities and generalizing the Möbius operator of two-dimensional conformal geometry \cite{calderbank_mobius_2006,burstall_conformal_2010}. This operator essentially corresponds to the possibility of defining (generalised) good-cut equations \cite{newman_heaven_1976,adamo_generalized_2010}. In the spirit of \cite{duval_carroll_2014} the data $\left(\scrI \to \gS, [n^{\mu}, h_{\mu\nu}],\cP\right)$ will be called strongly conformally Carrollian. As we will recall in our second section, strongly conformally Carrollian geometries are in one-to-one correspondence \cite{herfray_asymptotic_2020} with Cartan geometry modelled on \eqref{Introduction: Null infinity}. We will in fact recover these results from a more general (as compare to \cite{herfray_asymptotic_2020}) first order formalism that will encompass previous results and allows for a closer comparison with GHP-type formalisms \cite{geroch_spacetime_1973,frauendiener_new_2021,barnich_coadjoint_2021} and closely related Ehresmann connections \cite{campoleoni_two-dimensional_2019,ciambelli_carroll_2019,ciambelli_carrollian_2019,ciambelli_gauges_2020}.

 To further compare similitudes and differences between \eqref{Introduction: Carroll dS} and \eqref{Introduction: Null infinity}, it is instructive to introduce a third homogeneous Carrollian manifold with topology $\bbR \times S^2$:
\begin{equation}\label{Introduction: Null infinity in Bondi gauge1}
\scrI_{B}^{(3)} := \Quotient{\bbR^{4} \rtimes \SO\left(3\right)}{\bbR \times \ISO\left(2\right)}.
\end{equation}
The abelian factor in the denominator is a remaining Carrollian special conformal transformation; we will soon discuss its geometrical meaning. This model bridges between the two previous ones in the sense that we have the following maps
 \begin{equation*}
 \scrI^{(3)} \quad\to\quad \scrI^{(3)}_{B} \quad\to\quad Carr_{dS}^{(3)}.
 \end{equation*}
 The first map is rather natural: it is obtained by choosing an isometry subgroup $\SO\left(3\right)$ inside the conformal group $\SO\left(3,1\right)$ i.e. by picking up a round sphere representative in the conformal class of metric on $S^2$.  Accordingly $\scrI_{B}^{(3)}$ is a model for \emph{null infinity in Bondi gauge} (by which we really mean null infinity with a fixed constant curvature metric representative for the 2D metric). The second map is obtained by noting that 
 \begin{equation}\label{Introduction: Null infinity in Bondi gauge2}
 \scrI_{B}^{(3)} = \Quotient{\bbR \times \Carr_{dS}\left(3\right)}{\bbR \times \ISO\left(2\right)}
 \end{equation}
 and ``forgetting'' about the overall abelian factors. Note that, contrary to what the above expression might suggests, the $\bbR$ factor does act non trivially on the homogeneous space and there is no canonical way to perform this last step. The significance of this is the following: working in Bondi gauge, the geometry of the characteristic data at null infinity can be understood as a Carrollian geometry supplemented by an equivalence class $[\covD]$ of compatible connections \cite{ashtekar_radiative_1981,ashtekar_geometry_2015,ashtekar_null_2018} and the extra abelian factor in the denominator of \eqref{Introduction: Null infinity in Bondi gauge2} takes one representative on another $\nabla \mapsto \nabla + f h_{\rho\nu} n^{\mu}$, $f\in \Co{\scrI}$. Going from \eqref{Introduction: Null infinity in Bondi gauge2} to \eqref{Introduction: Carroll dS} amounts to choosing a preferred connection with constant curvature in the equivalence class. This sequence of gauge choices, leading from strongly conformally Carrollian to strong Carrollian geometries, is summarized in Table \ref{Table: introduction summary}. More details, especially on how these choices can be understood from the perspective of the corresponding Cartan geometries, will be given in the last section of this article.
 
	  \begin{table}
 	\caption{Summary}\label{Table: introduction summary}

 	\includegraphics[width= 1\linewidth]{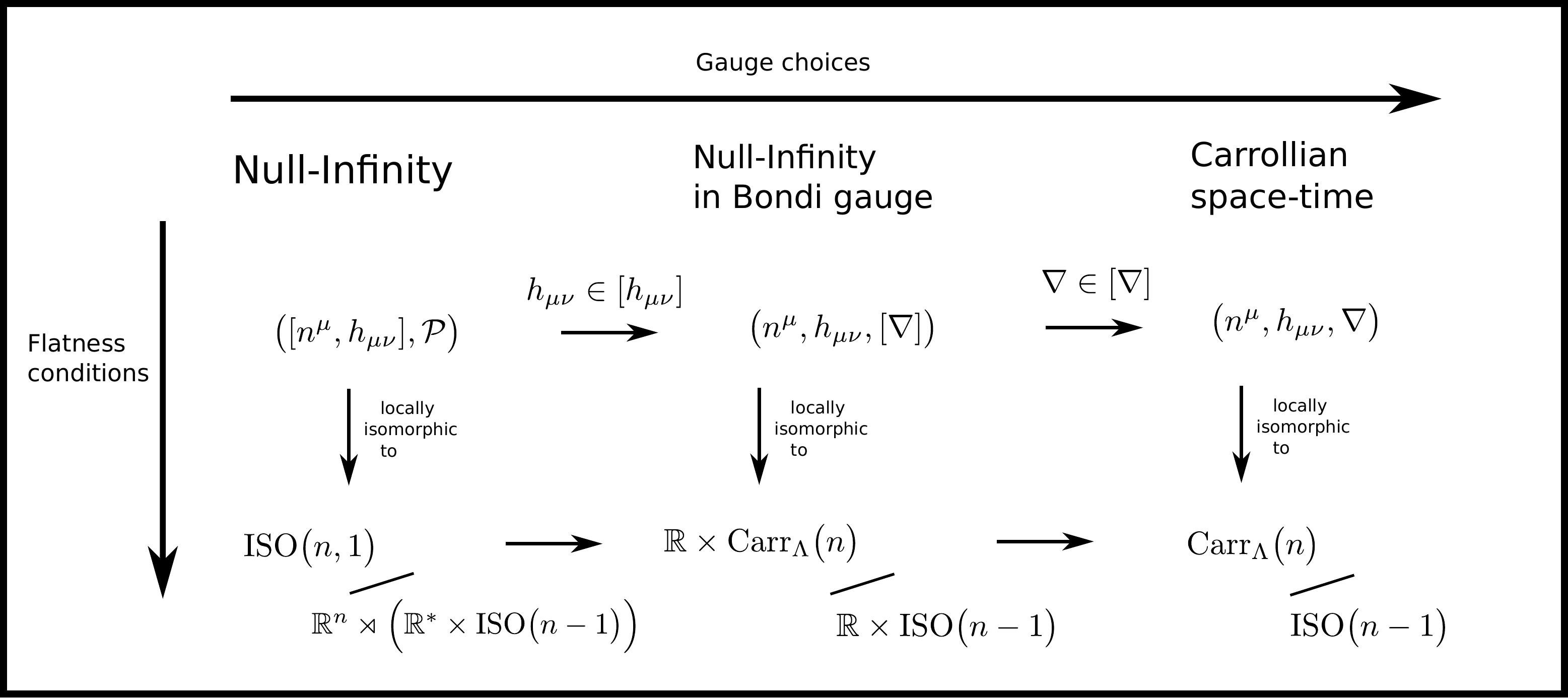}
 \end{table}

Let us close this introduction by emphasising the conceptual implications of these results for the physics of gravitational radiation. As we already discussed, characteristic data for gravity at null infinity precisely correspond to Cartan geometries modelled on \eqref{Introduction: Null infinity}, prior to any gauge fixing, or to other models \eqref{Introduction: Carroll dS},\eqref{Introduction: Null infinity in Bondi gauge2} after suitable choice of gauge. The curvature of the Cartan connection in fact matches \cite{herfray_tractor_2022} the Newman-Penrose coefficients $\Psi^0_4$, $\Psi^0_3$, $Im\left(\Psi^0_2\right)$ and therefore invariantly encodes the presence of gravitational radiation at null infinity. The significance of this fact comes from the fundamental theorem of Cartan geometry:
\begin{Theorem}{Fundamental theorem of Cartan geometry (see e.g. \cite{sharpe_differential_1997})}\label{Introduction: fundamental theorem}\\
	Let $\left(\cG \to M , \go\right)$ be a Cartan geometry\footnote{Let us recall to the reader that a Cartan geometry modelled on $G/H$ is the data of a $H$-principal bundle $\cG \to M$ together with a $\fkgg$-valued connection such that $\go \from T\cG \to \fkgg$ is invertible. Important examples are given by the projection $G \to G/H,$ together with the Maurer-Cartan connection $\go_{G}$ on $G$.} modelled on $G/H$, the curvature $d\go + \tfrac{1}{2}[\go,\go]$ vanishes if and only if the geometry is locally isomorphic to the model $\left(G \to G/H , \go_{G}\right)$. 
\end{Theorem}
In other terms, \emph{gravitational radiation is the obstruction to the existence of an isomorphism between null infinity and the model \eqref{Introduction: Null infinity}}. The use of Cartan geometry thus reveals the deep connection between gravitational waves and the BMS group: radiation at null infinity is the obstruction (in the precise sense of Theorem \ref{Introduction: fundamental theorem}) to being able to single out a preferred Poincaré group inside the BMS group. This well-known fact, which is of fundamental importance \cite{bondi_gravitational_1962,sachs_gravitational_1962,sachs_asymptotic_1962} here appears in a transparent form and has the value of a mathematical theorem. 

It also follows that in the absence of gravitational radiation the geometry of null infinity is equivalent to a flat Cartan connection, which by the fundamental theorem of Cartan geometry, is equivalent to an isomorphism from null infinity to the model. These are not unique and the space of all such isomorphisms physically correspond to inequivalent gravity vacua on which the BMS group act transitively \cite{compere_vacua_2016,ashtekar_null_2018,herfray_asymptotic_2020}. This degeneracy of gravity vacua is tightly related to the existence of memory effect in general relativity \cite{christodoulou_nonlinear_1991,bieri_gravitational_2015,ashtekar_geometry_2015,strominger_gravitational_2016} and has deep implications for the quantum gravity S-matrix \cite{strominger_bms_2014,strominger_lectures_2018,ashtekar_null_2018} as well as the black-hole information paradox \cite{hawking_soft_2016}. Finally, since gravity vacua correspond to flat Cartan connections on $\scrI$ it is natural to think of the corresponding Chern-Simon functional at null infinity \cite{nguyen_effective_2021} as some sort of effective boundary theory (see also \cite{herfray_einstein_2022} for more tractor actions for gravity). \\

This article is organised as follows: In a first section we  discuss the correspondence between strong Carrollian geometries and Cartan geometry modelled on \eqref{Introduction: Carroll dS}. In a second section we investigate the invariant geometry of null infinity as Cartan geometry modelled on \eqref{Introduction: Null infinity}. Finally we compare these two realisations by considering null infinity in Bondi gauge, i.e. for which one has singled out a preferred constant curvature representative from the conformal class.

\section{Carrollian spacetimes}

Let $\scrI$ be a manifold foliated by lines. This will be convenient (but not necessary) to suppose that the space of lines $\gS$ is a manifold and that consequently $\scrI \xto{\pi} \gS$ is a fibre bundle \footnote{This is a global requirement and can be overlooked if one is only interested in the local geometry}. We will also suppose that the lines are generated by a (nowhere vanishing) vector field $n^{\mu}$. The flow generated by this vector field equips (at least locally) Carrollian geometry with a preferred time $u \from U\subset \scrI \to \bbR$ which is unique up to ``super-translations'' 
\begin{equation}\label{Carrollian spacetimes: super-translations}
u \mapsto u + \pi^*\xi, \qquad \xi \in \Co{\gS}.
\end{equation}
\begin{Definition}\label{Carrollian spacetimes: Carrolian geometry def}
	A \emph{Carrollian geometry} \cite{duval_carroll_2014} $\left(\scrI \to \gS, n^{\mu}, h_{\mu\nu}\right)$ is the data of a one-dimensional fibre bundle $\scrI \to \gS$ together with a nowhere vanishing vector field $n^{\mu}$ generating the foliation and a degenerate metric $h_{\mu\nu}$ with one-dimensional kernel such that
	\begin{align}\label{Carrollian spacetimes: Lie derivative condition}
		h_{\mu\nu} n^{\nu} &=0, & \LieD_{n}h_{\mu\nu}&=0.
	\end{align}
\end{Definition}
The above requirements mean that $h_{\mu\nu}$ defines a genuine metric (i.e. invertible) on $\gS$. This definition has the advantage that Carrollian geometries then always have a non-trivial automorphism algebra given by the semi-direct product of super-translations $\cT \simeq \Co{\gS}$ with isometries of $\left(\gS, h_{\mu\nu}\right)$
\begin{equation}\label{Carrollian spacetimes: Carrollian symmetries}
\textrm{Lie}\left(\ISO\left(\scrI \to \gS, n^{\mu} , h_{\mu\nu} \right)\right) \simeq \textrm{Lie}\left(\cT \rtimes \ISO\left(\gS, h_{\mu\nu}\right)\right).
\end{equation}
This isomorphism is obtained by making a choice of local trivialisation (i.e. time). For instance, if $\left(\gS, h_{\mu\nu}\right) = \left( \bbR^{n-1} , dy^A dy^B \gd_{AB}\right)$ is flat euclidean space and $\scrI = \bbR \times \gS$ then symmetries are the semi-direct product of super-translations with euclidean isometries $\ISO\left(n-1\right) = \bbR^{n-1} \rtimes \SO\left(n-1\right)$.

Even though the second condition in \eqref{Carrollian spacetimes: Lie derivative condition} is too restrictive to encompass all geometries obtained by ultra-relativistic contractions, essential gravitational realisations are nonetheless of this type: prototypical examples are Killing horizons and null infinity in Bondi gauge. It should however be noted that the above definition does not encompass generic (non-Killing) null hypersurfaces for which there is no preferred vector field $n^{\mu}$, see \cite{ashtekar_isolated_2004} for a detailed discussion on alternative definitions modelling various type of horizons. A close inspection shows that none of the results in this section actually relies on the second requirement in \eqref{Carrollian spacetimes: Lie derivative condition}.

In this section we will discuss strongly Carrollian geometries \cite{duval_carroll_2014} (see \cite{morand_embedding_2020} for a comprehensive discussion) and describe the related Cartan geometries for a generic dimension and cosmological constant i.e. for Cartan geometries modelled on Carroll--De Sitter, Carroll and Carroll--Anti De Sitter. Taking the cosmological constant to zero we will recover results from \cite{hartong_gauging_2015}, restricting the arbitrary dimension to $2+1$ results of \cite{bergshoeff_three-dimensional_2017, ravera_ads_2019,matulich_limits_2019} and imposing a flatness condition those of \cite{figueroa-ofarrill_geometry_2019}. The main results are summarized in table \ref{Table: Carrolian geometry}.

\begin{table}
	 \caption{Cartan geometry of Carrollian spacetimes}\label{Table: Carrolian geometry}
\begin{center}
	\begin{tabular}{ccc}
	\hline\\
	\multicolumn{1}{l}{\textbf{Model}} &  &\\
	\multicolumn{3}{c}{
		\begin{tabular}{lcll}
			Carroll--De Sitter spacetime, $\gL >0$ & \hspace{0.5cm} & $\Quotient{\Carr_{dS}\left(n\right)}{\ISO\left(n-1\right)}$&$\simeq \bbR \times S^{n-1}$ \\ [1em]
			Carroll spacetime, $\gL =0$ && $\Quotient{\Carr\left(n\right)}{\ISO\left(n-1\right)}$  &$\simeq \bbR \times \bbR^{n-1}$  \\ [1em]
			Carroll--anti De Sitter spacetime, $\gL <0$ && $\Quotient{\Carr_{AdS}\left(n\right)}{\ISO\left(n-1\right)}$ &$\simeq \bbR \times H^{n-1}$\\ [1em]
		\end{tabular}
	} 
	\\[1em]	
	\hline\\
	\textbf{(Weak) Carrollian geometry} & &\\[1em]
	$\left(\scrI \to \gS, n^{\mu}, h_{\mu\nu}\right)$ & defines a canonical & $\ISO\left(n-1\right)$-principal bundle \\[0.4em]
	& has symmetry group & $\Co{\gS} \rtimes \ISO\left(\gS, h_{\mu\nu}\right)$ \\[1em] \hline\\
	\textbf{Strong Carrollian geometry} & &\\[1em]
	$\left(\scrI \to \gS, n^{\mu}, h_{\mu\nu}, \nabla \right)$& defines a unique & $\carr_{\gL}\left(n\right)$-valued connection $D$ \\ \\
	Flatness of $D$ &  \begin{tabular}{c}
		reduces the \\symmetry group to
	\end{tabular}  & $\Carr_{\gL}\left(n\right)$.\\[1em]
\hline
\end{tabular}
\end{center}
\end{table}

\subsection{Homogeneous model}\label{ss: Carrollian spacetimes, Homogeneous model}

The ``Carrollian isometry groups'' are obtained by ultra-relativistic contractions of the isometry group of de Sitter, flat, and anti de Sitter spacetimes in $n$ dimensions \cite{levy-leblond_nouvelle_1965,bacry_possible_1968}. The corresponding groups are respectively called Carroll--de Sitter, Carroll and Carroll--anti de Sitter. We will use the compact notation $Carr_{\gL}\left(n\right)$ for representing these three groups,
\begin{align*}
	\Carr_{dS}\left(n\right) &= \Carr_{\gL >0}\left(n\right),& \Carr\left(n\right) &= \Carr_{\gL =0}\left(n\right), & \Carr_{AdS}\left(n\right) &= \Carr_{\gL <0}\left(n\right).
\end{align*}
We then have
\begin{equation}\label{Carrollian spacetimes: Carroll Isometry Groups}
\Carr_{\gL}\left(n\right) := \bbR^n \rtimes \ISO_{\gL}\left(n-1\right)	
\end{equation}
where it is convenient to introduce a single notation
\begin{align}\label{Carrollian spacetimes: Riemannian Isometry Groups}
	\ISO_{\gL > 0}\left(n-1\right) &= \SO\left(n\right), & 	\ISO_{\gL=0}\left(n-1\right) &= \ISO\left(n-1\right),  &	\ISO_{\gL < 0}\left(n-1\right) &= \SO\left(n-1,1\right)_0.
\end{align}
for the (connected component of) isometry groups of $S^{n-1}$ , $\bbR^{n-1}$ and $H^{n-1}$. 

It is useful to think of $\ISO_{\gL}\left(n-1\right)$ as the subgroup of $\SO\left(n,1\right)_0$ stabilising a vector of norm $-2\gL$ (the factor of two is only introduced for future convenience). Making use of this fact, the Carrollian isometry groups can be represented matricially as
\begin{align}\label{Carrollian spacetimes: Carroll groups, matrix represantaton}
	\Carr_{\gL}\left(n\right)  = \left\{  \quad\begin{pmatrix} m^i{}_j & 0
		\\ t_i &1  \end{pmatrix} \quad \Big|\quad  \begin{array}{c} m^i{}_j \in \SO\left(n,1\right)_0,\; t_i \in \bbR^{n+1} \\[0.4em]  \text{s.t.} \quad  m^i{}_j I^j = I^i,\, t_i I^i=0. 
	\end{array} \quad\right\}
\end{align}
with $I^i \in \bbR^{n+1}$ a fixed vector of norm $-2\gL$. Introducing a null basis on $\bbR^{n+1}$ we write $I^i =\left(I^+, I^A , I^-\right)$ with $I^A \in \bbR^{n-1}$ and
\begin{equation*}
	I^i I^j \eta_{ij} = 2 I^+ I^- + I^A I^B \gd_{AB}.
\end{equation*}
Without loss of generality we can take $I^i= \left(1 , 0^A, -\gL\right)$. The Lie algebra $\carr_{\gL}\left(n\right)$ is then generated by matrices of the form
{\small \begin{align}\label{Carrollian spacetimes: Carroll Lie algebra}
\carr_{\gL}\left(n\right)	= \left\{  \quad\begin{pmatrix} 0 & -m_B & 0 &0
		\\ \gL\, m^A & \fkm^A{}_B &  m^A &0
		\\	0 & -\gL\, m_B & 0 &0
		\\ \gL\, t_- & t_B &  t_- &0 \end{pmatrix}  \quad \Big|\quad \begin{array}{c} \fkm^A{}_{B} \in \so(n-1),\;  t_B \in \bbR^{n-1}, \\[0.4em]  (m^A, t_-) \in \bbR^{n} 
		\end{array}  \quad\right\}.
\end{align}}\\
Spatial rotations and ``Carrollian boosts''  are respectively generated by $\fkm^A{}_B$ and $t_B$, while the $n$ ``Carrollian translations'' are generated by $m^A$ and $t_-$. In this representation, the Carroll isometry groups naturally act on $\bbR^{n+2}$ with coordinates $Y^I = \left( Y^+ , Y^A , Y^- , Y^0\right)$ and degenerate metric
\begin{equation}\label{Carrollian spacetimes: ambient space degenerate metric}
Y^I Y^J h_{IJ} = 2 Y^+ Y^- + Y^{A}Y^{B} \gd_{AB}.
\end{equation}

The homogeneous Carrollian spacetimes are then defined to be the quotient \cite{bacry_possible_1968,figueroa-ofarrill_spatially_2019,figueroa-ofarrill_geometry_2019}
\begin{equation}\label{Carrollian spacetimes: homogeneous space realisation}
	Carr^{(n)}_{\gL} := \Quotient{\Carr_{\gL}\left(n\right)}{\ISO\left(n-1\right)}. 
\end{equation}
Where $\ISO\left(n-1\right)$ is identified with the subgroup of $\Carr_{\gL}\left(n\right)$ stabilising $X^I = \left( 0, 0^A, 1, 0\right)$,
{\small \begin{align}\label{Carrollian spacetimes: Carroll stabilisator Group}
\left\{ \begin{pmatrix}1& 0 & 0 & 0\\ 0 & m^A{}_B & 0& 0 \\ 0 &  0 & 1 & 0 \\
0 & t_B & 0 & 1 \end{pmatrix} \quad \Big|\quad m^A{}_B \in \SO\left(n-1\right), t_B\in \bbR^{n-1} \quad\right\} &\leq \Carr_{\gL}\left(n\right),
\end{align}}
and therefore is generated by Carrollian boosts and rotations.

These models are the total space of a (trivial) line bundle over sphere/flat/hyperbolic spaces respectively (the degenerate metric is obtained by pull-back of the metric on the base) i.e. we have the isomorphisms
\begin{align*}
Carr^{(n)}_{dS}  &\simeq \bbR \times S^{n-1}, & Carr^{(n)}  &\simeq \bbR \times  \bbR^{n-1}, & Carr^{(n)}_{AdS}  &\simeq \bbR \times H^{n-1}.
\end{align*}

Alternatively, the model Carrollian spacetimes can be realised as
{\small \begin{equation}\label{Carrollian spacetimes: ambient space realisation}
Carr^{(n)}_{\gL} = \left\{Y^I = \Mtx{Y^+ \\ Y^A \\ Y^- \\ Y^0} \in \bbR^{n+2} \quad\big|\quad Y^I Y^J h_{IJ} =0,\quad  Y^I I^J_{(\gL)} h_{IJ} = 1   \right\}
\end{equation}}
\\
where $I^I_{(\gL)} := \left( 1, 0^A, -\gL, 0\right)$ is a fixed vector of norm $-2\gL$ for the degenerate metric \eqref{Carrollian spacetimes: ambient space degenerate metric} (the Carrollian vector field is then proportional to $\parD_{Y^0}$). In particular, note that $X^I = \left( 0, 0^A, 1, 0\right)$ always belongs to $Carr^{(n)}_{\gL}$. In order to identify \eqref{Carrollian spacetimes: ambient space realisation} with the homogeneous space \eqref{Carrollian spacetimes: homogeneous space realisation} it suffices to note that the Carrollian isometry groups act transitively and that the stabiliser of a point $X^I$ is $\ISO\left(n-1\right)$.

\subsection{Tetrad formalism adapted to Carrollian geometry}

Let $\left(\scrI \to \gS,n^{\mu}, h_{\mu\nu}\right)$ be a Carrollian geometry as in Definition \ref{Carrollian spacetimes: Carrolian geometry def}. Let $U$ be an open set of $\scrI$ and $\big\{e_A{}^{\mu} \big\}_{A\in 1,\,...\,,n-1}$, be vector fields on this set. We will say that $\left\{e_a{}^{\mu}\right\}_{a\in 1,\,...\,,n}= \left\{e_A{}^{\mu}, n^{\mu}\right\}_{A\in 1,\,...\,,n-1}$ is an orthogonal frame\footnote{The distribution spanned by the $\{e_A{}^{\mu} \}_{A\in 1,\,...\,,n-1}$ defines a connection on the line bundle $\scrI \to \gS$ in the sense of Ehresmann, see \cite{ciambelli_carroll_2019} for a discussion. Also note that working with such tetrads essentially amounts to Newman-Penrose formalism \cite{newman_approach_1962,newman_spin-coefficient_2009} or related GHP-type formalisms \cite{geroch_spacetime_1973,frauendiener_new_2021,barnich_coadjoint_2021}.} or ``tetrad''\footnote{The term ``tetrad'' only make literal sense for $n=4$ but we will here use it as generic term for local frames. } if it forms a basis of the tangent bundle at any $x \in U$ and satisfies
\begin{align*}
	e_A{}^{\mu} e_B{}^{\nu} h_{\mu\nu} &= \gd_{AB}
\end{align*}
where $\gd_{AB}$ is the identity matrix. Tetrads have a $\ISO\left(n-1\right)$-ambiguity
\begin{align}\label{Carrollian spacetimes: tetrad transformation rules}
	\begin{array}{lcl}
	e_A{}^{\mu} &\mapsto& m_A{}^B \left( e_B{}^{\mu}-  t_B n^{\mu} \right)\\
	n^{\mu} & \mapsto& n^{\mu}
	\end{array}&&\Mtx{m^A{}_B & 0 \\ t_B &1}&\in \ISO\left(n-1\right).
\end{align}
Carrollian geometries are therefore canonically equipped with a $\ISO\left(n-1\right)$-principal bundle obtained from the reduction of the frame bundle of $\scrI$ to such tetrads. These local symmetries are respectively interpreted as spatial rotations (for those generated by $\SO\left(n-1\right)$) and ``Carrollian boosts'' (generated by $\bbR^{n-1}$).

We will write the corresponding dual tetrad $\left\{ \gth_{\mu}{}^a \right\}_{a\in 1,\,...\,,n} =  \left\{ \gth_{\mu}{}^A , l_{\mu} \right\}_{A\in 1,\,...\,,n-1}$:
\begin{align*}
	e_B{}^{\mu}\gth_{\mu}{}^A &= \gd_B{}^A, & n^{\mu}\gth_{\mu}{}^A &= 0, & e_B{}^{\mu}l_{\mu} &= 0, & n^{\mu}l_{\mu}&=1.
\end{align*}
In what follows, it will be important that, under the action of $\ISO\left(n-1\right)$, dual tetrads transform as
\begin{align}\label{Carrollian spacetimes: dual-tetrad transformation rules}
\gth_{\mu}{}^A &\mapsto m^A{}_B\; \gth_{\mu}{}^B, & l_{\mu} & \mapsto l_{\mu}+ t_C\; \gth_{\mu}{}^C.
\end{align}

\subsection{Strongly Carrollian geometry}

Let $\left( \scrI \to \gS, n^{\mu}, h_{\mu\nu}\right)$ be a Carrollian geometry. We will say that it is strongly Carrollian \cite{duval_carroll_2014,morand_embedding_2020} if, on top of this, it is equipped with a torsion-free connection $\covD_{\gr}$ on the tangent bundle satisfying the compatibility relations
\begin{align}\label{Carrollian spacetimes: compatibility relations}
	\covD_{\gr} n^\mu &=0, & \covD_{\gr} h_{\mu\nu} & =0.
\end{align}
The components of this connection in a tetrad $\left\{ e_A{}^{\mu}, n^{\mu}\right\}$ are
\begin{align}\label{Carrollian spacetimes: affine connection coordinates}
\Mtx{ \gth_{\mu}{}^A \covD_{\gr} e_B^{\mu} &  \gth_{\mu}{}^A \covD_{\gr} n^{\mu} \\ l_{\mu} \covD_{\gr} e_B^{\mu} & l_{\mu} \covD_{\gr} n^{\mu}  } =\Mtx{\go_{\gr}{}^A{}_B & 0 \\ -\tfrac{1}{2}C_B{}_{\gr} & 0}
\end{align}
(the minus one half factor is here added to fit with notation from the literature \cite{barnich_aspects_2010,herfray_tractor_2022}) and the torsion-free condition is
\begin{align*}
	d^{\covD}\left(\gth^{a} e_a{}^{\mu}\right):=d^{\go}\gth^A\; e_A^{\mu} + \left(dl -\tfrac{1}{2} C{}_B \W \gth^B \right) n^{\mu}=0
\end{align*}
where we introduced the notation $d^{\go}\gth^A := d\gth^A + \go^A{}_B \W \gth^B$. It follows that $\go_{\gr}{}^A{}_B$ is uniquely determined in terms of $\gth^A_{\mu}$. On the other hand, the coefficients\footnote{Note that, even though we order tetrads with the degenerate element coming in last position, we write the corresponding coefficient with a zero. }  \begin{align*}
C_{AB} \;\gth_{\mu}{}^B + C_{A0} \, l_{\mu} &:= C_A{}_{\mu}
\end{align*}
are only fixed in terms of the tetrad up to
\begin{equation*}
C_A{}_{\mu} \mapsto C_A{}_{\mu} +  \gd C_{AB} \;\gth_{\mu}{}^B
\end{equation*}
where $\gd C_{AB}$ is a symmetric tensor, $\gd C_{AB} = \gd C_{(AB)}$. This remaining freedom corresponds to the information contained in the strongly Carrollian geometry $\left( \scrI \to \gS, n^{\mu}, h_{\mu\nu}, \covD \right)$ as opposed to the bare Carrollian geometry $\left( \scrI \to \gS, n^{\mu}, h_{\mu\nu}\right)$.

\subsection{The Carrollian Cartan connection}

Carrollian spacetimes are tightly related to the geometry of $\carr_{\gL}$-valued connections (see e.g. \cite{hartong_gauging_2015,bergshoeff_three-dimensional_2017,matulich_limits_2019,figueroa-ofarrill_geometry_2019}). 

Let $\left(\scrI \to \gS, n^{\mu}, h_{\mu\nu}\right)$ be a Carrollian geometry. Corresponding tetrads then form an $\ISO\left(n-1\right)$-principal bundle and we will call (Carrollian) tractors (following the terminology of \cite{bailey_thomass_1994,cap_parabolic_2009}) sections $\Phi^I$ of the associated bundle for the representation \eqref{Carrollian spacetimes: Carroll stabilisator Group}.

 Let $D_{\gr}$ be a generic connection valued in the Carroll Lie algebra \eqref{Carrollian spacetimes: Carroll Lie algebra}, it acts on tractors as
\begin{equation*}
	D\Phi^I = \Mtx{
		d & -\gth_B  & 0 & 0 \\
		\gL \, \gth^A & d^{\go} &  \gth^A &0 \\
		0 & -\gL\, \gth_B & d & 0 \\
			\gL\, l & -\tfrac{1}{2}C_B &  l &d 
	}\Mtx{\Phi^+ \\ \Phi^B \\ \Phi^- \\ \Phi^0}.
\end{equation*} 
Changes of tetrad \eqref{Carrollian spacetimes: tetrad transformation rules} are equivalent to gauge transformations with values in $\ISO\left(n-1\right)$ (in the representation \eqref{Carrollian spacetimes: Carroll stabilisator Group}) and induce the following transformation on the connection components
\begin{align*}
	\gth^A &\mapsto m^A{}_B\gth^B,  & l &\mapsto l + \gth^C t_C,
\end{align*}
\begin{align}\label{Carrollian spacetimes: gauge transformation of fields - C}
	\go^A{}_B &\mapsto m^A{}_C \go^C{}_D m_B{}^D - dm^A{}_C m_B{}^C,&	-\tfrac{1}{2}C_A & \mapsto m_A{}^B\left(-\tfrac{1}{2}C_B -d^{\go}t_B\right).	
\end{align}
We will say that such a connection is compatible with the Carrollian geometry if $\left\{\gth_{\mu}{}^A , l_{\mu} \right\}$ defines a dual-tetrad. The curvature of the connection is
\begin{equation*}
	F^I{}_J = \Mtx{0 & -d^{\go} \gth_B  &0 &0 \\
		\gL \, d^{\go} \gth^A & F^A{}_B - 2 \gL\, \gth^A \W \gth_B  & d^{\go}\gth^A &0\\
		0 & - \gL\, d^{\go} \gth_B & 0 &0	\\
			\gL \left( dl -\tfrac{1}{2} C_C \W \gth^{C}\right) & -\tfrac{1}{2}d^{\go} C_B - 2 \gL\, l \W \gth_B  &dl -\tfrac{1}{2} C_C \W \gth^C &0}.
\end{equation*}
In particular $\left(\go^A{}_B, C_B\right)$ are the components of a torsion-free connection $\covD$ if and only if $F^I{}_-=0$. We will say that such $\carr_{\gL}\left(n\right)$-valued connections are torsion-free.

We thus derived the following, generalizing results from \cite{hartong_gauging_2015}:
\begin{Proposition}\label{Carrolian spacetimes: proposition, Carrolian connection}
	Let $\left(\scrI \to \gS, n^{\mu}, h_{\mu\nu}\right)$ be a $n$-dimensional Carrollian geometry, there is a one-to-one correspondence between, on the one hand, $\carr\left(n\right)$-valued connections which are both compatible with the Carroll geometry and torsion-free and, on the other hand, choice of strongly Carrollian geometries $\left(\scrI \to \gS, n^{\mu}, h_{\mu\nu}, \covD\right)$.
\end{Proposition}
It follows from the fundamental theorem of Cartan geometry (see theorem \ref{Introduction: fundamental theorem}) that we have the 
\begin{Proposition}
	Let $\left(\scrI \to \gS, n^{\mu}, h_{\mu\nu}, \covD\right)$ be a $n$-dimensional strongly Carrollian geometry satisfying
	\begin{align*}
	F^A{}_B - 2 \gL\; \gth^A \W \gth_B &=0,	&  -\tfrac{1}{2}d^{\go} C_B - 2 \gL\; l \W  \gth_B &=0,
	\end{align*}
then $\scrI \to \gS$ is locally isomorphic to the model 
\begin{equation*}
Carr^{(n)}_{\gL} := \Quotient{\Carr_{\gL}\left(n\right)}{\ISO\left(n-1\right)}. 
\end{equation*}
In particular, the algebra of local symmetries is $\carr_{\gL}\left(n\right)$.
\end{Proposition}

\subsection{``Carrollian good-cuts''}

\subsubsection{BMS coordinates}

We will call ``BMS'' coordinates $ u \from \scrI \to \bbR$ a local trivialisation\footnote{Strictly speaking the trivialisation $\scrI \simeq \bbR \times \gS$ is given by $ x \mapsto(u(x) , \pi(x))$.} of $\scrI \to \gS$ chosen such that $n^{\mu} (du)_{\mu} =1$ i.e. such that it defines an element of dual tetrad $l_{\mu} := (du)_{\mu}$. This defines a unique tetrad $\{e_A{}^{\mu}, n^{\mu}\}$ (up to $\SO\left(n-1\right)$) by requiring that $e_A^{\mu} (du)_{\mu}=0$. Changing the trivialisation $u \mapsto \uh$ gives the following transformation rules
\begin{align*}
e_A{}^{\mu} &\mapsto e_A{}^{\mu}-  t_A n^{\mu}, & t_B &= \covD_{B}\uh,
\end{align*}
where $\covD_A\uh := e_A^{\mu}\covD_{\mu}\uh$.

In BMS coordinates, $dl=0$ and the previous computation simplifies, for example normal Carrollian connections are then parametrised by $C_A = C_{(AB)}\gth^B$ and $\go^A{}_B$ is just the pull-back of the Levi-Civita connection of $\left(\gS, h_{\mu\nu}\right)$ to $\scrI$. 

\subsubsection{Carrollian good-cuts}

Let $\cG \from \gS \to \scrI$ be a section (or ``cut'') of $\scrI$. Let $u$ be a BMS coordinate. We can always think of the cut $\cG$ as the zero set of another BMS coordinate $f$ on $\scrI$,
\begin{equation*}
f = u - \pi^* G
\end{equation*}
where $G\in \Co{\gS}$ is a function on $\gS$ and the cut is given by $f=0$.

 We will say that $\cG$ is a Carrollian good-cut for a strongly Carrollian geometry $\left(\scrI\to \gS, n^{\mu}, h_{\mu\nu}, \covD_{\gr}\right)$ if and only if
\begin{equation*}
\covD_{(\mu} \covD_{\nu)} f \big|_{\cG}=0.
\end{equation*}
In a tetrad given by $u$ this is equivalent to
\begin{equation*}
\covD_{(A} \covD_{B)} G - \tfrac{1}{2}C_{AB}\big|_{u=G} = 0.
\end{equation*}
(this follows from \eqref{Carrollian spacetimes: affine connection coordinates}). In the flat model \eqref{Carrollian spacetimes: homogeneous space realisation} the space of Carrollian good-cut is $n$ dimensional and defines preferred time foliations on which the $n$ Carrollian translations act transitively.

These ``Carrollian good-cuts'' crucially differ from the usual good-cut equation \cite{newman_heaven_1976,hansen_r._o._metric_1978,adamo_generalized_2010,adamo_null_2012} of null-infinity\footnote{Strictly speaking Newman's good-cut equations are obtained for $n=3$ and considering only the holomorphic part of this: One can always write the 2D metric as $h = P(z,\zb)^{-2} 2dz d \zb$ and we then have $(P\parD_z)^A(P\parD_z)^B\left(\covD_A \covD_B G\right) = \parD_z \left( P^2 \parD_z G\right) = \eth^2 G$. Introducing $\gs := (P\parD_z)^A(P\parD_z)^B \tfrac{1}{2}C_{AB}$ one recovers the canonical form $\eth^2 G - \gs\big|_{u=G}=0$ for the (generalised \cite{adamo_generalized_2010}) good-cut equations.}
\begin{equation*}
\left(\covD_{(A} \covD_{B)} G - \tfrac{1}{2}C_{AB}\big|_{u=G}\right)\big|_{tf} = 0.
\end{equation*}
 by a trace part (here $\big|_{tf}$ indicates ``trace-free part of''). While the usual good-cut equation has at most $(n+1)$ independent solutions, the Carrollian good-cuts must satisfy one more equation and have a maximum of $n$ independent solutions. This is ultimately related to the fact that the homogenous realisation of Carrollian geometry \eqref{Carrollian spacetimes: homogeneous space realisation} has $n$ Carrollian translations while the homogenous realisation of null-infinity \eqref{Null infinity: homogeneous space realisation} possess $n+1$ of these. This can be clarified by relating good-cuts to covariantly constant sections.

\subsubsection{Covariantly constant sections}

Let $\left(\scrI \to \gS, n^{\mu}, h_{\mu\nu}, \covD \right)$ be a strongly Carrollian geometry and let $D$ be the corresponding Carroll-valued connection. Its action on dual elements is
\begin{equation*}
D\Phi_I = \Mtx{
	d & -\gth^B  & 0 & -l \\
	\gL \gth_A& d^{\go} &  \gth_A & \tfrac{1}{2}C_A \\
	0 & -\gL \gth^B & d & - l\gL \\
	0 & 0 &  0 &d 
}\Mtx{\Phi_- \\ \Phi_B \\ \Phi_+ \\ \Phi_0}
\end{equation*} 
where duality is defined by $\Phi_I \Phi^I = \Phi_+\Phi^+ + \Phi_A\Phi^A + \Phi_-\Phi^- + \Phi_0\Phi^0$. By construction the Carrollian connection stabilises the section $I_{(\gL)}^I = \left( 1 , 0^A , -\gL , 0\right)$.

Let $\Phi_I = \left(f , \Phi_A, \Phi_+ , \Phi_0\right)$ be a covariantly constant dual tractor $D_{\gr}\Phi_I=0$ satisfying $ \Phi_+ - \gL f =\Phi_I I^I_{(\gL)}= 0$. A direct computation shows that this is equivalent to having $\Phi_I = \left( f, \covD_A f, \gL f, \fd  \right)$ and
\begin{align}\label{Carrollian spacetimes: covariantly constant tractor equations}
\covD_{\mu} \fd &=0, & \covD_{(A} \covD_{B)} f + \tfrac{1}{2}C_{(AB)} \fd + 2\gL f \; h_{AB} &= 0
\end{align}
where dot indicate Lie derivatives along $n^{\mu}$.

Let us pick a BMS coordinate $u \from \scrI \to \bbR$. The above equations can be equivalently rewritten as
\begin{align}\label{Carrollian spacetimes: covariantly constant tractor equations in BMS coordinates}
f &= k \left(u - G\right)  , & \covD_{(A} \covD_{B)} G - \tfrac{1}{2}C_{AB} - 2\gL \left(u -G\right)\; h_{AB} &= 0,
\end{align}
where $G$ is a function on $\gS$ and $k\in \bbR$ is some constant. Evaluating the above equation at $u=G$ one finds that $\Phi_I$ is covariantly constant if and only if $G$ solves the Carrollian good cut equation and 
\begin{align*}
	\tfrac{1}{2}C_{AB} = - 2\gL (u-G) h_{AB} + \tfrac{1}{2}C_{AB}\big|_{u=G}.
\end{align*}
I.e. $\tfrac{1}{2}\Cd_{AB} = - 2\gL h_{AB}$. This last condition amounts to the vanishing of one of the coefficient of the Carrollian curvature. When this is satisfied Carrollian good-cuts are in one-to-one correspondence with covariantly constant sections $\Phi_I$ which are orthogonal to $I_{\gL}^I$ and considered up to an overall scale. This is a space of finite dimension at most $n$ and this dimension is attained when the geometry is flat.

\section{Null Infinity}

Since null infinity is the conformal boundary of an asymptotically flat spacetime this must be some type of conformal manifold. This is in fact a conformal Carrollian geometry in the following sense.
\begin{Definition}\label{Null infinity: Conformal Carroll geometry def}
	A \emph{conformal Carrollian geometry} $\left(\scrI \to \gS , [n^{\mu}, h_{\mu\nu}]\right)$ is the data of a $n$-dimensional fibre bundle $\scrI \to \gS$ over a $(n-1)$-dimensional manifold $\gS$ together with an equivalence class of vector fields and degenerate metrics with one-dimensional kernel
	\begin{equation*}
	\left(n^{\mu}, h_{\mu\nu}\right) \sim \left( \gl^{-1} n^{\mu},\gl^{2} h_{\mu\nu}\right), \quad   \qquad \Co{\scrI} \ni \gl >0,
	\end{equation*}
	satisfying
	\begin{align*}
	n^{\nu}h_{\mu\nu} &=0, & \LieD_{n}h_{\mu\nu}&= \gTh h_{\mu\nu},
	\end{align*}
where\footnote{Here $h^{\mu\nu}$ is any inverse of $h_{\mu\nu}$.} $\gTh := \tfrac{1}{n-1}h^{\mu\nu}\LieD_{n}h_{\mu\nu}$.
\end{Definition}
The above definition means that $\left(\gS ,[h_{\mu\nu}]\right)$ defines a conformal geometry in the usual sense. We stress, however, that in principle rescalings by functions $\gl$ defined on the whole of $\scrI$ are allowed. As is well known \cite{duval_conformal_2014,ashtekar_geometry_2015}, the algebra of automorphisms of conformal Carrollian geometry is a generalised version of the BMS algebra.
\begin{equation}\label{Null infinity: BMS symmetries}
\textrm{Lie}\left(\ISO\left(\scrI \to \gS, [n^{\mu} , h_{\mu\nu}] \right)\right) \simeq \textrm{Lie}\left(\cT \rtimes \ISO\left(\gS, [h_{\mu\nu}]\right)\right),
\end{equation}
We recover the standard BMS group \cite{bondi_gravitational_1962,sachs_gravitational_1962,sachs_asymptotic_1962} by considering the case where $\left(\gS , [h_{\mu\nu}]\right)$ is the two-dimensional conformal sphere, $\textrm{BMS}_4 \simeq \cT \rtimes \SO\left(3,1\right)$. Note that choosing a representative $\left(\scrI \to \gS , n^{\mu}, h_{\mu\nu}\right)$ such that $\LieD_{n}h_{\mu\nu}=0$ is always possible (these representatives are Carrollian geometries in the sense of Definition \ref{Carrollian spacetimes: Carrolian geometry def}), we will call these ``Bondi gauge'' and discuss them in the next section.

Conformal geometry is typically a difficult subject because of the necessity of dealing with Weyl rescalings. Tractor calculus \cite{bailey_thomass_1994,curry_introduction_2018} has been devised specifically to provide a manifestly ``Weyl-rescaling invariant'' formalism. In \cite{herfray_asymptotic_2020} these methods were extended to conformal Carrollian manifolds. We here keep developing this formalism, making use of a more general first order (tetrad-like) approach which will encompass all previous results. This allows for a closer comparison with GHP-type formalisms \cite{geroch_spacetime_1973,frauendiener_new_2021,barnich_coadjoint_2021} or (closely related) Ehresmann connections \cite{campoleoni_two-dimensional_2019,ciambelli_carroll_2019,ciambelli_carrollian_2019,ciambelli_gauges_2020}. We emphasise that the approach presented here is in principle far more powerful since it is manifestly Weyl invariant by construction (without requiring extra structure such as Weyl connections etc).

 The main difference between the Carrollian and the classical conformal situation is that a conformally Carrollian geometry is not associated with a unique normal Cartan connection (as opposed to the situation in standard conformal geometry, see e.g. \cite{curry_introduction_2018}). Rather, there is a family of possible connections corresponding to choices of strongly conformal Carrollian geometries (these results are summarised in Table \ref{Table: Null infinity}). 

From the point of view of null infinity, this moduli space of Cartan connections correspond to choices of gravitational characteristic data; a non-vanishing curvature then characterises the presence of gravitational radiation \cite{herfray_tractor_2022}. 

It is finally worth noticing that it has in fact long been known \cite{penrose_twistor_1973} that, in dimension $d=4$, the local twistor connection restricted to null infinity invariantly encodes the asymptotic shear of an asymptotically flat spacetime. However, it seems that the definition of the tractor bundle for conformally Carrollian geometry, the corresponding Cartan geometries and their relations to spacetime local-twistors have only been systematically investigated for the first time in \cite{herfray_asymptotic_2020,herfray_tractor_2022}.

\begin{table}
	\caption{Cartan geometry of null infinity}\label{Table: Null infinity}
\begin{center}
\begin{tabular}{p{4cm}cc}
	\hline\\
	\multicolumn{1}{l}{\textbf{Model}} &  &\\[0.5em]
	\multicolumn{3}{c}{  $\scrI^{(n)}:=$ $\qquad \Quotient{\ISO\left(n,1\right)}{\bbR^{n} \rtimes \left( \bbR^* \times \ISO\left(n-1\right)\right)} \quad \simeq\quad  \bbR \times S^{n-1}$} \\ [1.5em]
	
	\hline\\
	\multicolumn{3}{l}{\textbf{(Weak) conformally Carrollian geometry}} \\[1.5em]
	$\left(\scrI \to \gS, [n^{\mu}, h_{\mu\nu}] \right)$ & defines a canonical & {\small $\bbR^{n} \rtimes \left( \bbR^* \times \ISO\left(n-1\right)\right)$}-principal bundle \\[1em]
	& has symmetry group & $\Co{\gS} \rtimes \ISO\left(\gS, [h_{\mu\nu}]\right)$ \\[1em] \hline\\
	\multicolumn{3}{l}{\textbf{Strong conformally Carrollian geometry}}\\[1.5em]
	$\left(\scrI \to \gS, [n^{\mu}, h_{\mu\nu}], \cP \right)$& defines a unique & $\iso\left(n,1\right)$-valued connection $D$ \\ \\
	\begin{tabular}{l}
		Flatness of $D$ \end{tabular} &  \begin{tabular}{c}
		reduces the \\symmetry group to
	\end{tabular}  & $\ISO\left(n,1\right)$.\\[1em]
\hline
\end{tabular}
\end{center}
\end{table}

\subsection{Homogeneous model}

The homogeneous model for null infinity is
\begin{equation}\label{Null infinity: homogeneous space realisation}
\scrI^{(n)} := \Quotient{\ISO\left(n,1\right)}{\bbR^{n}\rtimes\left(\bbR^*\times \ISO\left(n-1\right)\right)}.
\end{equation}
It is the total space of a fibre bundle over the conformal sphere
\begin{equation*}
\scrI^{(n)} \simeq \bbR \times S^{n-1}.
\end{equation*}

In a sense, the homogeneous realisation of null infinity \eqref{Null infinity: homogeneous space realisation} is a conformal generalisation of the Carrollian spacetimes \eqref{Carrollian spacetimes: homogeneous space realisation}. This is obvious at the level of the isometry group since the Carroll groups can all be understood as subgroups of the Poincaré group. It suffices to compare the matrix representation \eqref{Carrollian spacetimes: Carroll groups, matrix represantaton} of $\Carr_{\gL}\left(n\right)$ with
\begin{align*}
\ISO\left(n,1\right) = \left \{ \quad \begin{pmatrix} m^i{}_j & 0
\\ t_i &1  \end{pmatrix} \quad \Big| \quad \begin{array}{c} m^i{}_j \in \SO\left(n,1\right)_0,\; t_i \in \bbR^{n+1}
\end{array} \quad \right\}.
\end{align*}
In particular, in the basis for $\bbR^{n+2}$ introduced in section \ref{ss: Carrollian spacetimes, Homogeneous model}, the Lie algebra of the $(n+1)$-dimensional Poincaré group is generated by matrices of the form
{\small \begin{align}\label{Null infinty: Poincaré Lie algebra}
 \iso \left(n,1\right) = \left\{  \begin{pmatrix} \ga & -m_B & 0 &0
\\ r^A & \fkm^A{}_B &  m^A &0
\\	0 & -r_B & -\ga&0
\\ t_+ & t_B &  t_- &0 \end{pmatrix} \quad \Big|\quad \begin{array}{c} \fkm^A{}_{B} \in \so(n-1),\;  t_B \in \bbR^{n-1},\, \ga \in \bbR,  \\[0.4em]   (m^A, t_-) \in \bbR^{n+1}, \, (r^A, t_+) \in \bbR^{n+1} 
\end{array}  \quad    \right \} ,
\end{align}}{}\\
On top of the spatial rotations, Carrollian boosts and translations respectively generated by $\fkm^A{}_B$, $t_B$ and $m^A$, $t_-$ we have, as compare to Carrollian isometries, extra dilatation generated by $\ga$ and $n$ extra Carrollian special conformal transformations generated by $r^A$ and $t_+$.

The subgroup of $\ISO\left(n,1\right)$ stabilising the line generated by $X^I = \left(0 , 0^A , 1 , 0\right)$ is isomorphic to $\bbR^{n} \rtimes \left(\bbR^* \times\ISO\left(n-1\right)\right)$ and is realised as
{\small \begin{align}\label{Null infinity: Poincaré stabilisator Group}
\Bigg\{\begin{pmatrix} \gl & 0 & 0 & 0\\ r^A & m^A{}_B & 0& 0 \\ -\gl^{-1}\tfrac{1}{2} r^2 &  -\gl^{-1}r_C m^C{}_B & \gl^{-1} & 0 \\
t_+ & t_B & 0 & 1 \end{pmatrix}, \begin{tabular}{c} $m^A{}_B \in \SO\left(n-1\right),\; r_B\in \bbR^{n-1}$ \\[0.3em] $\gl\in \bbR^*,\; t_+ \in \bbR,\; t_B \in \bbR^{n-1}$ \end{tabular} \Bigg\} &\leq \ISO\left(n,1\right).
\end{align}}{}\\
Here spatial rotations $m^A{}_B$ and Carrollian boosts $r_B$, which were already present in the Carrollian model \eqref{Carrollian spacetimes: Carroll stabilisator Group}, are complemented by dilatations $\gl$ to form a semi-direct product with respect to Carrollian special conformal transformations $(r^A, t_+) \in \bbR^{n}$. The homogeneous model for null infinity is then obtained as the quotient \eqref{Null infinity: homogeneous space realisation}.

Alternatively, the model for null infinity can be realised as the space of null lines in $\bbR^{n+2}$
{\small \begin{equation}\label{Null infinity: ambient space realisation}
\scrI^{(n)} = \left\{Y^I = \sMtx{Y^+ \\ Y^A \\ Y^- \\ Y^0} \in \bbR P^{(n+1)} \quad\big|\quad  Y^I Y^J h_{IJ} = 0   \right\}
\end{equation}}\mbox{}\\
where the square bracket indicates homogeneous coordinates on the projective space $\bbR P^{(n+1)}$ and $h_{IJ}$ is the degenerate metric \eqref{Carrollian spacetimes: ambient space degenerate metric}. The line generated by $X^I = \left(0 , 0^A , 1 , 0\right)$ is null and therefore defines a point of $\scrI^{(n)}$. The identification of \eqref{Null infinity: ambient space realisation} with \eqref{Null infinity: homogeneous space realisation} is obtained by noting that the Poincaré group acts transitively on \eqref{Null infinity: ambient space realisation} and that the stabiliser of a point $X^I$ is $\bbR^{n} \rtimes \left(\bbR^* \times\ISO\left(n-1\right)\right)$.

Comparing this construction with the one from section \ref{ss: Carrollian spacetimes, Homogeneous model}, it is clear that one obtains the homogeneous Carrollian spacetimes from the model for null infinity by introducing an extra vector $I^I_{(\gL)} =\left(1, 0^A, -\gL, 0\right)$. This extra structure breaks conformal invariance and reduces the isometry group from Poincaré to the corresponding Carroll group.

\subsection{Conformal geometry of null infinity}

\subsubsection{Bundle of scale}

If $\scrI$ is orientable we define the bundle of scale $L\to \scrI$ as $L := \big(\gL^{n}T^*\scrI\big)^{-\tfrac{1}{n}}$ (if $\scrI$ is not orientable one needs to use densities instead but this won't be relevant here). 

Let $\left(\scrI \to \gS , [n^{\mu}, h_{\mu\nu}]\right)$ be a conformal Carrollian geometry. Let $\left\{e_A{}^{\mu}, n^{\mu}\right\}$ be a choice of tetrad and $\left\{\gth_{\mu}{}^A, l_{\mu}\right\}$ the related dual tetrad, in particular it defines preferred representatives
\begin{equation*}
\left(n^{\mu}, \gd_{CD}\; \gth_{\mu}{}^C \;\gth_{\nu}{}^D\right) \in  [n^{\mu}, h_{\mu\nu}]
\end{equation*} 
and a preferred volume form which can be used to trivialise the bundle of scale: any section $\bf$ of $L^k$ can be rewritten as
{\small \begin{equation*}
 \bf = f	\left(l \W \gth^1 \W ... \W \gth^{n-1} \right)^{-\tfrac{k}{n}}
\end{equation*}}
where $f$ is a function on $\scrI$. Upon rescaling the tetrad
\begin{align}\label{Null infinity: tetrad transformation rules}
	\begin{array}{lcl}
		e_A{}^{\mu} &\mapsto& \gl^{-1} \; m_A{}^B\left( \;e_B{}^{\mu}-  t_B n^{\mu}\right)\\
		n^{\mu} & \mapsto& \gl^{-1}\; n^{\mu}
	\end{array}&&\gl\Mtx{m^A{}_B & 0 \\ t_B &1}&\in  \bbR^* \times \ISO\left(n-1\right),
\end{align}
  we have the transformations rules for the dual tetrad
 \begin{align}\label{Null infinity: dual-tetrad transformation rules}
 \gth_{\mu}{}^A &\mapsto \gl m^A{}_B\; \gth_{\mu}{}^B, & l_{\mu} & \mapsto \gl\left(l_{\mu}+ t_C\; \gth_{\mu}{}^C\right),
 \end{align} 
  and therefore the transformation rules for (coordinates representatives of) sections of $L^k$,
\begin{align}\label{Null infinity: weighted function transformation rules}
	\left(h_{\mu\nu} , n^{\mu}\right) &\mapsto \left( \gl^2 h_{\mu\nu} , \gl^{-1} n^{\mu}\right), & f \mapsto \gl^k f.
\end{align}
This gives a practical definition of sections of $L^k$. In particular, one could rephrase the definition of conformal Carroll geometries as a choice $\left(\bh_{\mu\nu} , \bn^{\mu}\right)$ where $\bh_{\mu\nu}$ and $\bn^{\mu}$ respectively are valued in sections of $L^2$ and $L^{-1}$ respectively. We will typically abuse notation and identify coordinate representatives $f$ (following transformation rules \eqref{Null infinity: weighted function transformation rules}) with the corresponding section $\bf$ of $L^k$.

\paragraph{Vertical connection}

In presence of a conformal Carroll structure $\left(\scrI \to \gS, [n^{\mu},h_{\mu\nu}] \right)$ there is a natural differential operator on the bundle of scale $\covD_0^{(\gTh)} \from \So{L} \to \Co{\scrI}$ defined as follows. Let $\left\{\gth_{\mu}{}^A, l_{\mu}\right\}$ be a choice of dual tetrad and let us introduce the quantity $\gTh$ defined as
{\small \begin{align*}
	\gTh &:= \frac{2}{n-1} \frac{\LieD_{n}\left(\gth^{1} \W ... \W \gth^{n-1}\right) }{\gth^{1} \W ... \W \gth^{n-1}} \\ &=\frac{1}{n-1} h^{\mu\nu} \LieD_{n} h_{\mu\nu}.
\end{align*}}
Under the change of tetrad \eqref{Null infinity: tetrad transformation rules} we have
{\small \begin{equation*}
	\gTh \mapsto \frac{1}{\gl}\left(\gTh +2\tfrac{\gld}{\gl}\right)
\end{equation*}}
where $\gld :=\LieD_{n}\gl$. If $f$ is a section of $L$ we now define the ``vertical derivative'' as
\begin{equation*}
	\covD_{0}^{(\gTh)}f := \fd - \tfrac{1}{2}\gTh f.
\end{equation*}
It is invariant under the change of tetrad \eqref{Null infinity: tetrad transformation rules} and therefore defines a function on $\scrI$. We will say that $f \in \So{L}$ has constant vertical derivative if $\covD_{0}^{(\gTh)}f$ is a constant on $\scrI$ i.e.
\begin{equation}\label{Geometry of Null infinity: constant vertical derivative}
	\covD_{\mu}\left(\covD_{0}^{(\gTh)}f\right) =0.
\end{equation}

\subsubsection{Strongly conformally Carrollian geometry}\label{sss: Strongly conformally Carrollian geometry}

Let $\left(\scrI\to \gS, [n^{\mu}, h_{\mu\nu}]\right)$ be a conformally Carrollian geometry and let $\Quotient{S^2 T^*\scrI}{h}$ be the bundle obtained by taking the quotient of symmetric tensors by $h_{\mu\nu}$,
\begin{equation*}
	T_{\mu\nu} \sim T_{\mu\nu} + \ga h_{\mu\nu}, \qquad \ga \in \Co{\scrI}.
\end{equation*}
We want to define \emph{Poincaré operators} $\cP$ as second order differential operators 
\begin{equation*}
	\cP \from \So{L} \to \So{\Quotient{S^2 T^*\scrI}{h} \otimes L}.
\end{equation*}
To do so let us take a dual tetrad $\left(\gth^A, l\right)$ and let $\go^A{}_B$ be the one form obtained by solving $d^{\go} \gth^A=0$. We define the covariant derivative\footnote{This is a useful notation but potentially misleading, note in particular that if $\covD_{\mu}$ is a Carrollian connection as in \eqref{Carrollian spacetimes: affine connection coordinates} then $e_A{}^{\mu} e_B{}^{\nu} \left( \covD_{\mu} V_{\nu}\right) = \covD_{A}V_{B} + \tfrac{1}{2}C_{BA} V_0$.} as $\covD_{A} V_B  := e_{A}{}^{\mu}\left( \parD_{\mu}V_{B} - \go^C_{\mu}{}_B V_{C}\right)$. It will also be useful to introduce the notations $(\LieD_{n}l)_A := e_A{}^{\mu}(\LieD_{n}l)_{\mu}$, $(\LieD_{n}\gth^A)_B := e_B{}^{\mu}(\LieD_{n}\gth^A)_{\mu}$ etc.

Let $f$ be a section of $L$ i.e. transforming as $f \mapsto \gl f$ under \eqref{Null infinity: tetrad transformation rules} and let us consider
\begin{subequations}\label{Null infinity: Poincaré operator}
\begin{equation}\label{Null infinity: Poincaré operator only}
	\cP\left(f\right)_{\mu\nu} = \cP_{00}(f)\; l_{\mu}\; l_{\nu} +  \cP_{A0}(f) \; 2l_{(\mu}\; \gth^A_{\nu)} + \cP(f)_{AB} \; \gth^A_{\mu}\; \gth^B_{\nu}.
\end{equation}
with
\begin{align}
	&\cP_{00}(f)  = \covD_{0} \covD_{0}{}^{(\gTh)}f, \\
	&\cP_{A0}(f)  =\covD_{A}\covD_{0}{}^{(\gTh)}f,\\
	&\cP\left(f\right)_{AB}= \covD_{(A}\covD_{B)} \big|_{tf}\; f  +\tfrac{1}{2}C_{AB} \fd - \Big( \tfrac{1}{2} \left(\LieD_{n}C\right)_{AB} -\covD_{(A} (\LieD_{n}l)_{B)} + (\LieD_{n}l)_{A}(\LieD_{n}l)_{B} \Big)\big|_{tf} \; f
\end{align}
\end{subequations}
where $\big|_{tf}$ indicates ``trace-free part'', dot indicates Lie derivative along $n$ and $C = C_{AB}\gth^A\gth^B$ is symmetric and trace-free. We stress that $C$ is here arbitrary and parametrizes the Poincaré operator. Comparing with \eqref{Geometry of Null infinity: constant vertical derivative} one sees that the above generalises the ``constant vertical derivative operator'' and one can also think of this operator as a conformally invariant modification of $\eqref{Carrollian spacetimes: covariantly constant tractor equations}$.

Equation \eqref{Null infinity: Poincaré operator} defines a section of $\Quotient{S^2 T^*\scrI}{h} \otimes L$, i.e. transforms as 
\begin{equation*}
\cP\left(f\right)_{\mu\nu} \mapsto \gl \cP\left( f\right)_{\mu\nu} + \ga h_{\mu\nu},  \qquad \ga \in \Co{\scrI},
\end{equation*}
under the change of tetrad \eqref{Null infinity: tetrad transformation rules}, if and only if we take the transformation rules for $C_{AB}$ to be
\begin{equation}\label{Null infinity: C transformation rules}
\tfrac{1}{2}C_{AB} \mapsto \gl^{-1} m_A{}^C m_B{}^D\left( \tfrac{1}{2}C_{CD} + \covD_{D}t_C + \left( \left(\tfrac{\gld}{\gl} + \tfrac{\gTh}{2}\right)t_C- \gl^{-1}\covD_C\gl   - (\LieD_{\nh} \lh )_C \right)t_D  \right)
\end{equation}
(here $\nh$, $\lh$ are the images of \eqref{Null infinity: tetrad transformation rules}, \eqref{Null infinity: dual-tetrad transformation rules}). This results from a tedious but direct computation, we will however see that it straightforwardly follows from tractor calculus adapted to conformally Carrollian geometry.

With the transformation rules \eqref{Null infinity: C transformation rules} the expression \eqref{Null infinity: Poincaré operator} therefore defines a differential operator $\cP$ taking a section of $L$ to a section of the quotient bundle $\Quotient{S^2 T^*\scrI}{h} \otimes L$. These operators play in conformal Carrollian geometry a similar role to the Möbius operators \cite{calderbank_mobius_2006,burstall_conformal_2010} of two-dimensional conformal geometry and for this reason were called Poincaré operators in \cite{herfray_asymptotic_2020}.

Let $\left( \scrI \to \gS, [n^{\mu}, h_{\mu\nu}]\right)$ be a conformal Carrollian geometry. We will say that it is a \emph{strong conformally Carrollian geometry} \cite{herfray_asymptotic_2020} if, on top of this, it is equipped with choice of Poincaré operator \eqref{Null infinity: Poincaré operator} i.e. a choice of trace-free ``tensor'' $C_{AB}$ transforming as \eqref{Null infinity: C transformation rules}.

The main interest of this definition is that the conformal boundary of an asymptotically flat spacetime is always strongly conformally Carrollian \cite{herfray_tractor_2022}. This is because the asymptotic shear $C_{AB}$ of asymptotically flat spacetimes follows the transformation rules \eqref{Null infinity: C transformation rules} (this can also be obtained more invariantly by the use of tractor methods). As we will discuss in details later on, strongly conformally Carrollian geometries are also equivalent \cite{herfray_asymptotic_2020} to Cartan geometries modelled on \eqref{Null infinity: homogeneous space realisation}.

\subsection{The null-tractor bundle}\label{ss: The null-tractor bundle}

 \subsubsection{Null-tractors: invariant definition}\label{sss: Null-tractors: invariant definition}

Let $J^2 L$ be the second order jet of sections of $L$ on $\scrI$ and let $\left(a , a_{\mu}, a_{\mu\nu}\right)$ be corresponding coordinates. Let us consider the sub-bundle $F\subset J^2 L$ of formal solutions to \eqref{Geometry of Null infinity: constant vertical derivative}.
Practically, $F$ is the sub-bundle of $J^2 L$ given by the condition 
\begin{equation*}
n^{\nu}a_{\mu\nu} +\tfrac{1}{2} \gTh \;a_{\mu} +\tfrac{1}{2} \covD_{\mu}\gTh \;a =0.	
\end{equation*}
There is always a canonical inclusion $S^2T^*\scrI \otimes L \to J^2 L$ 
\begin{equation*}
	a_{\mu\nu} \to \left(0, 0, a_{\mu\nu}\right),
\end{equation*}
and, requiring $n^{\mu}a_{\mu\nu}=0$, this gives a canonical inclusion $S^2 (T\scrI /n)^* \otimes L \to F$. The dual null-tractor bundle on $\scrI$ is then defined as the quotient of $F$ by the image of symmetric trace-free tensors in $F\subset J^2 L$
\begin{equation}\label{Null infinity: null-tractor bundle}
	\cT^* := \Quotient{F}{	S^2\big|_{tf} (T\scrI /n)^* \otimes L}.
\end{equation}
This definition is useful because it is invariant. It clearly shows that any conformal Carrollian geometry has a canonical tractor bundle. However it is not very practical which is the reason why we now move to a more concrete realisation.

\subsubsection{Null-tractor: transformation rules}

Let $\gs\in \So{L}$ be a section with constant vertical derivative, i.e. satisfying \eqref{Geometry of Null infinity: constant vertical derivative}. For such sections, and after making a choice of tetrad, we define Thomas operator $T\from \So{L} \xto{e} \So{\bbR \oplus \bbR^{n-1}\oplus \bbR \oplus \bbR}$ as:
{\small \begin{equation}\label{Null Infinity: Thomas operator}
\begin{array}{ccc}
\gs & \xto{e} & \Mtx{\Phi_- \\ \Phi_A \\ \Phi_+ \\ \Phi_0} := \Mtx{ \gs \\ \covD_A \gs \\ -\tfrac{1}{n-1}\left(\covD^C \covD_C - P\right)\gs \\ \covD^{(\gTh)}_0 \gs}
\end{array}	
\end{equation}}{}\\
where, for $n\geq3$, $P =\tfrac{1}{2(n-2)}R$ is the trace of the Schouten tensor of $\covD_A$. (Recall that if $\left\{\gth^A, l\right\}$ is a dual tetrad and $\go^A{}_B$ is the one form obtained by solving $d^{\go} \gth^A=0$, we define the covariant derivative $\covD_A V_{B}$ as $\covD_{A} V_B  := e_{A}{}^{\mu}\left( \parD_{\mu}V_{B} - \go^C_{\mu}{}_B V_{C}\right)$).

Thomas operator depends on a choice of tetrad. Upon the change of tetrad \eqref{Null infinity: tetrad transformation rules} we have the transformation rules
{\small \begin{equation*}
\Mtx{\Phi_- \\ \Phi_A \\ \Phi_+ \\ \Phi_0} \mapsto \Mtx{ \gl & 0 & 0& 0\\ r_A & m_A{}^B & 0 & -m_A{}^C t_C \\ -\gl^{-1}\tfrac{1}{2}r^2 & - \gl^{-1} r^C m_C {}^B & \gl^{-1}& \gl^{-1}\left(r^D m_D{}^C t_C - t_+\right) \\ 0&0&0&1} \Mtx{\Phi_- \\ \Phi_B \\ \Phi_+ \\ \Phi_0}
\end{equation*}}
where $r_A$, $t_+$ satisfy
\begin{align}\label{Null Infinity: tractor transformation rules r_A t_+}
r_A & = m_A{}^C\left(\gU_C -\left( \tfrac{\gld}{\gl} + \tfrac{\gTh}{2}\right)t_C\right), & t_+ &= \tfrac{1}{n-1}\left( -\covD_C t^C + \tfrac{1}{2}\LieD_{n}(t^2) + t^C (\LieD_{n}l)_C\right),
\end{align}
where $\gU_C = \lambda^{-1} \covD_A \lambda$. We can now \emph{define} dual tractors as fields following the above transformation rules.

From the transformation behaviour of dual tractors, we obtain transformation rules for tractors
\begin{equation}\label{Null Infinity: tractor transformation rules}
\Mtx{\Phi^+ \\ \Phi^A \\ \Phi^- \\ \Phi^0} \mapsto \Mtx{\gl & 0 & 0& 0\\ r^A & m^A{}_B & 0 & 0 \\ -\gl^{-1}\tfrac{1}{2}r^2 & - \gl^{-1}r_C m^C{}_B& \gl^{-1}& 0 \\ t_+&t_B&0&1} \Mtx{\Phi^+ \\ \Phi^B \\ \Phi^- \\ \Phi^0}
\end{equation}
with $r_A$, $t_+$ required to satisfy \eqref{Null Infinity: tractor transformation rules r_A t_+}. Here the pairing between tractors and their dual is taken to be $\Phi^I \Phi_I := \Phi^+ \Phi_+ + \Phi^A \Phi_A + \Phi^-\Phi_- + \Phi^0 \Phi_0$. In the spirit of this subsection we can \emph{define} tractors as fields following the above transformations rules.

\subsubsection{Splitting isomorphisms}

We presented two definitions of (null-)tractors: the first is invariant but unpractical, the second is very concrete but hides the underlying geometrical content. These are in fact equivalent and having the two is a good thing.

The equivalence of the two definitions can be seen as follows: if $\{e_A{}^{\mu}\}$ is a choice of tetrad, the Thomas operator (given by equation \eqref{Null Infinity: Thomas operator}) defines an isomorphism
\begin{equation}\label{Null infinity: splitting isomorphism}
\begin{array}{ccc}
\cT^* & \xto{e} & \bbR \oplus \bbR^{n-1} \oplus \bbR \oplus \bbR
\end{array}
\end{equation}
between the dual-tractor bundle \eqref{Null infinity: null-tractor bundle} and $\bbR^{n+2}$. By duality we also have a similar isomorphism for tractors. We will refer to such isomorphisms, one for each choice of tetrad, as \emph{splitting isomorphisms}. If $\Phi^I$ is a section of the tractor bundle, we will write
{\small \begin{equation}\label{Null Infinity: splitting isomorphism}
\Phi^I \xeq{e} \Mtx{\Phi^+ \\ \Phi^A \\ \Phi^- \\ \Phi^0}.
\end{equation}}
Changing the tetrad
\begin{align*}
\begin{array}{lcl}
e_A{}^{\mu} &\mapsto & \eh_A{}^{\mu}:= \gl^{-1} \; m_A{}^B\left( \;e_B{}^{\mu}-  t_B n^{\mu}\right)\\
n^{\mu} & \mapsto &\nh := \gl^{-1}\; n^{\mu}
\end{array}&&\gl\Mtx{m^A{}_B & 0 \\ t_B &1}&\in \ISO\left(n-1\right) \times \bbR,
\end{align*}
then induces a change of splitting isomorphism 
{\small \begin{equation*}
\Phi^I \xeq{e} \Mtx{\Phi^+ \\ \Phi^A \\ \Phi^- \\ \Phi^0} \quad \mapsto \quad \Phi^I \xeq{\eh} \Mtx{\Phih^+ \\ \Phih^A \\ \Phih^- \\ \Phih^0}
\end{equation*}}
corresponding to the transformation rules \eqref{Null Infinity: tractor transformation rules}.

\subsubsection{\texorpdfstring{The $\bbR^{n} \rtimes \left(\bbR^* \times \ISO\left(n-1\right)\right)$-principal bundle of null infinity}{The R x (R x ISO(n-1))-principal bundle of null infinity}}

Comparing the transformation rules of tractors \eqref{Null Infinity: tractor transformation rules} with \eqref{Null infinity: Poincaré stabilisator Group} one sees that null-tractors can in fact be thought as associated to a  $\bbR^{n-1} \rtimes \left(\bbR^* \times \ISO\left(n-1\right)\right)$-principal bundle over $\scrI$. This principal bundle is the bundle of ``orthonormal frames'' for null-tractors.

It indeed follows from the transformation rules \eqref{Null Infinity: tractor transformation rules} that the null-tractor bundle is equipped with an invariant degenerate pairing $h_{IJ}$, a preferred degenerate section $I^I$ and a preferred null weighted section $X^I \in \So{\cT \otimes L}$ 
\begin{align}\label{Null infinity: tractor metric}
h_{IJ} \Phi^I \Phi^J &\xeq{e} 2\Phi^+\Phi^- + \Phi^A\Phi^B \gd_{AB}, & I^I &\xeq{e} \Mtx{0 \\ 0 \\ 0 \\1}, & X^I &\xeq{e} \Mtx{0\\0\\1\\0}. 
\end{align}
It suffices to check that these are invariant under the transformation rules \eqref{Null Infinity: tractor transformation rules}.

We now define orthonormal tractor frames $\{ e_{+}{}^I, e_A{}^I , e_-{}^I, I^I \}$ as tractor frames satisfying
\begin{align*}
e_-{}^I &\propto X^I,&	e_+{}^{I} e_-{}{}^J h_{IJ} &=1, & e_A{}^{I} e_B{}^J h_{IJ} &= \gd_{AB},
\end{align*}
with all other contractions vanishing. These frames form a $\bbR^n \rtimes\left( \bbR \times \ISO\left(n-1\right)\right)$-principal bundle over $\scrI$.

One concludes that conformal Carrollian geometries $\left(\scrI \to \gS, [n^{\mu}, h_{\mu\nu}]\right)$ are canonically equipped with a $\bbR^n \rtimes\left( \bbR \times \ISO\left(n-1\right)\right)$-principal bundle. We stress that this bundle is as intrinsic to conformal Carrollian geometry as the $\ISO\left(n-1\right)$-principal bundle of frames \eqref{Carrollian spacetimes: tetrad transformation rules} was to Carrollian geometry.

\subsubsection{Spacetime interpretation}

It was shown in \cite{herfray_tractor_2022} that if $\left(\scrI \to \gS, n^{\mu}, h_{\mu\nu}\right)$ is the conformal boundary of an $(n+1)$-dimensional asymptotically flat spacetime then the null-tractor bundle \eqref{Null infinity: null-tractor bundle} can be canonically identified with a sub-bundle of the restriction at $\scrI$ of the standard tractor bundle \cite{bailey_thomass_1994} of the $(n+1)$-dimensional spacetime. 

A choice of representative $g_{\mu\nu} \in[g_{\mu\nu}]$ for the conformal spacetime metric then defines a splitting isomorphism \eqref{Null infinity: splitting isomorphism} for the tractor bundle. The Weyl parameter $\go\left(r\right)$ of a smooth Weyl rescaling $g_{\mu\nu} \mapsto \go^2 g_{\mu\nu}$ must have an expansion of the form
\begin{equation*}
\go\left(r\right) = \go_{0} +  r^{-1} \go_{1} + \cO\left(r^{-2}\right)
\end{equation*}
and the tractor transformation rules \eqref{Null Infinity: tractor transformation rules} then receive the following interpretation: $\gl = \go_0$ corresponds to Weyl rescalings of the boundary degenerate metric while $t_+ =\go_1$ corresponds to the action of sub-leading Weyl rescalings. Finally, the restriction \eqref{Null Infinity: tractor transformation rules r_A t_+} correspond to restricting to representatives $g_{\mu\nu}$ satisfying the BMS gauge $\parD_{r} \textrm{det}\left(h_{AB}\right)=0$.

\subsection{The tractor connection}

\subsubsection{\texorpdfstring{The $\ISO\left(n,1\right)$-valued connection}{The ISO(n,1)-valued connection}}

Let $\left(\scrI\to \gS , [n^{\mu}, h_{\mu\nu}]\right)$ be a conformal Carrollian geometry, recall that the corresponding tractor bundle is equipped with a degenerate metric $h_{IJ}$ and degenerate direction $I^I$ given by \eqref{Null infinity: tractor metric}. Let $D_{\gr}$ be a connection on the null-tractor bundle satisfying $D_{\gr}h_{IJ}=0$, $DI^I=0$.  In a given tetrad, it must be of the form
\begin{equation*}
D\Phi^I \xeq{e} \Mtx{
	d +\ga & -\gth_B  & 0 & 0 \\
	-\xi^A& d^{\go} &  \gth^A &0 \\
	0 & \xi_B & d -\ga & 0 \\
	-\psi & -\tfrac{1}{2}C_B &  l &d 
}\Mtx{\Phi^+ \\ \Phi^A \\ \Phi^- \\ \Phi^0}.
\end{equation*} 
Note that this amounts to requiring that $D_{\gr}$ takes values in the Lie algebra of the Poincaré group \eqref{Null infinty: Poincaré Lie algebra}.

The action of the $\bbR^{n-1} \rtimes \left(\bbR \times\ISO\left(n-1\right)\right)$-valued gauge transformations \eqref{Null Infinity: tractor transformation rules} on the connection is given by
\begin{subequations}\label{Null Infinity: gauge transformation of fields}
\begin{align*}
\gth^A &\mapsto \gl m^A{}_B\gth^B,  & l &\mapsto \gl\left(l + \gth^C t _C\right),
\end{align*}
\begin{align}\label{Null Infinity: gauge transformation of fields - ga}
\ga &\mapsto \ga + r^C m_C{}^B \gth_B - \gl^{-1}d\gl,
\end{align}
\begin{align*}
\go^A{}_B &\mapsto m^A{}_C \go^C{}_D m_B{}^D - dm^A{}_C m_B{}^C + \left(m^A{}_C r_B - r^A m_{BC}\right) \gth^C,
\end{align*}
\begin{align*}
	\xi^A &\mapsto \gl^{-1} m^A{}_B\left(\xi^B + d_{\go-\ga} \left(r^C m_C{}^B\right)\right) + \gl^{-1} \left(\tfrac{1}{2}r^2 \gd^A{}_B - r^A r_B\right)m^B{}_C \gth^C,
\end{align*}
\begin{align}\label{Null Infinity: gauge transformation of fields - C}
	-\tfrac{1}{2}C_A & \mapsto m_A{}^B\left( -\tfrac{1}{2}C_B -d^{\go}t_B - \gth_B t_+\right) + r_A \left(l +\gth^C t_C \right),
\end{align}
\begin{align*}
\psi& \mapsto \gl^{-1}\left(\psi + d_{\ga}t_+ + \xi^C t_C - r^C m_C{}^B \left( \tfrac{1}{2}C_B + d_{\go}t_B +\gth_B t_+\right) + \tfrac{1}{2} r^2 \left( l + \gth^B t_B\right) \right).
\end{align*}
\end{subequations}
Just as in the Carrollian case, we will say that such a connection is compatible with the Carrollian geometry if $\{\gth_{\mu}{}^A, l_{\mu}\}$ defines a null tetrad for $\left(h_{\mu\nu}, n^{\mu}\right)$.

Importantly the tractor transformation rules are such that $r_A$ and $t_+$ must satisfy \eqref{Null Infinity: tractor transformation rules r_A t_+}. A direct computation shows that it implies that we in fact have
\begin{align}\label{Null infinity: gauge fixing condition on the connection}
	\ga + \tfrac{\gTh}{2}l & \mapsto \ga + \tfrac{\gTh}{2}l, & 
	\tfrac{1}{2} h^{CD} C_{CD} & \mapsto \gl^{-1} \tfrac{1}{2} h^{CD} C_{CD} + \gl^{-1}\left( (\LieD_{n}l)_C -\tfrac{1}{2}C_{C0} \right)t^C
\end{align}
where $\gTh := \tfrac{1}{n-2}h^{CD}\hd_{AB}$, $C_{A} = C_{AB} \gth^B + C_{A0} l$. We will mainly consider torsion-free connections, these must satisfy $ \tfrac{1}{2}C_{C0}- (\LieD_{n}l)_C=0$ and therefore for such connections we can invariantly require that
\begin{align}\label{Null infinity: compatibility with Thomas operator}
	\ga + \tfrac{\gTh}{2}l &=0, & 
	\tfrac{1}{2} h^{CD} C_{CD} & =0.
\end{align}

We stress that we can only impose \eqref{Null infinity: compatibility with Thomas operator} because the transformation rules for tractors satisfy \eqref{Null Infinity: tractor transformation rules r_A t_+}. Since these particular transformation rules are given by Thomas operator, we will say that the connection is compatible with Thomas operator if and only if it satisfies \eqref{Null infinity: compatibility with Thomas operator}. Alternatively, depending on the reader's mindset, one can take the transformation rules \eqref{Null Infinity: tractor transformation rules r_A t_+} to be defined as the subset of \eqref{Null Infinity: gauge transformation of fields - ga}, \eqref{Null Infinity: gauge transformation of fields - C} preserving the ``gauge fixing conditions'' \eqref{Null infinity: compatibility with Thomas operator} for torsion-free connections (more generally, they are the transformation rules such that \eqref{Null infinity: gauge fixing condition on the connection} hold). Even though valuable, we wish to highlight that this point of view might hide the important fact that the null-tractor bundle and its preferred transformation rules \eqref{Null Infinity: tractor transformation rules} with \eqref{Null Infinity: tractor transformation rules r_A t_+}  are canonically given by the conformal Carrollian geometry.

The curvature of the connection is
\begin{equation}\label{Null infinity: curvature}
F^I{}_J \xeq{e} \Mtx{ d\ga-\xi_C\W \gth^C & -d_{\go+\ga} \gth_B  &0 &0 \\
	-d_{\go-\ga} \xi^A & F^A_{\go}{}_B +\xi^A \W \gth_B +\gth^A\W \xi_B  & d_{\go+\ga}\gth^A &0\\
	0 & d_{\go-\ga} \xi_B & -d\ga +\xi_C \W \gth^C &0	\\
	-d_{-\ga}\psi +\tfrac{1}{2}  C_C \W \xi^{C} & -\tfrac{1}{2}d^{\go} C_B + \psi \W \gth_B + l \W \xi_B &d_{\ga}l -\tfrac{1}{2}C_C \W \gth^C &0}.
\end{equation}
Let us write
\begin{align*}
\xi_A &= \xi_{AB} \gth^B + \xi_{A0}l& C_{A} &= C_{AB} \gth^B + C_{A0}l.
\end{align*}
The requirement that the connection is torsion-free i.e. satisfies $F^I{}_J X^J=0$ uniquely fixes $ \xi_{A0}$, $C_{A0}$ and the skew symmetric parts $\xi_{[AB]}$, $C_{[AB]}$:
\begin{align*}
	\go_0{}^A{}_B &= \tfrac{\gTh}{2}\gd^A{}_B - e_{B}{}^{\mu}(\LieD_{n}\gth^A)_{\mu}, & \go_C^A{}_B & = -e_C{}^{\mu}e_B{}^{\nu} \parD_{\mu} \gth^A_{\nu},\\
	\tfrac{1}{2} C_{A0} &= e_{A}{}^{\mu}(\LieD_{n}l)_{\mu},& C_{[AB]}&= - e_A{}^{\mu}e_B{}^{\nu} (dl)_{\mu\nu},\\
	\xi_{A0} &= e_A{}^{\mu}\left(\tfrac{1}{2}\covD_{\mu} \gTh +\tfrac{\gTh}{2} (\LieD_{n}l)_{\mu}\right), & \xi_{[AB]} &= \tfrac{\gTh}{4} e_A{}^{\mu}e_B{}^{\nu} (dl)_{\mu\nu}.
\end{align*}

One needs further ``normality conditions'' (these generalise the normality condition of conformal geometry \cite{bailey_thomass_1994,sharpe_differential_1997,cap_parabolic_2009}) to constrain the components of $\psi = \psi_0 l + \psi_B \gth^B$ and the symmetric part $\xi_{(AB)}$ in terms of the other fields. In this context a set of rather natural invariant normality conditions are
\begin{subequations}
\begin{align}
n\intD \left(F^A_{\go}{}_B +\xi^A \W \gth_B +\gth^A\W \xi_B\right) &=0,\\
n \intD \left(-\tfrac{1}{2}d^{\go} C_B + \psi \W \gth_B + l \W \xi_B\right)&=0, \label{Null Infinity: normality conditions1} \\
e^B\intD \left(F^A_{\go}{}_B +\xi^A \W \gth_B +\gth^A\W \xi_B\right) &=0, \label{Null Infinity: normality conditions2}\\
e^B\intD \left(-\tfrac{1}{2}d^{\go} C_B + \psi \W \gth_B + l \W \xi_B\right) &=0.\label{Null Infinity: normality conditions3}
\end{align}
\end{subequations}
We will say that connections satisfying the above constraints are ``normal.'' In fact the first equation above, necessary for the  third to be invariant under gauge transformations, is identically satisfied and does not add any further constraint on the field. Supposing that compatibility conditions \eqref{Null infinity: compatibility with Thomas operator} hold, equation \eqref{Null Infinity: normality conditions1} is equivalent to
\begin{align}
	\xi_{AB}\Big|_{tf} + \tfrac{\gTh}{4} C_{AB} &=  \Big( \tfrac{1}{2} \LieD_{n}\left(C\right)_{AB}  + \covD_{A} (\LieD_{n}l)_{B} + (\LieD_{n}l)_{A} \; (\LieD_{n} l)_{B}\Big)\Big|_{tf},\label{Null Infinity: normality conditions1'}\\
	(n-1)\psi_0 &= -\xi^C{}_C - \covD^C (\LieD_{n}l)_{C} +  (\LieD_{n}l)_{C} \; (\LieD_{n}l)^C, \nonumber
\end{align}
where it is again convenient to write $(\LieD_{n}l)_C := e_C{}^{\mu}(\LieD_{n}l)_{\mu}$ etc and $C= C_{AB}\gth^A\gth^B$. These equations fix $\psi_0$ and the trace-free symmetric tensor $\xi_{(AB)}\big|_{tf}$ (one can check that \eqref{Null Infinity: normality conditions1'} does not impose any further constraint on the skew symmetric part). Equations \eqref{Null Infinity: normality conditions2} and \eqref{Null Infinity: normality conditions3} are respectively equivalent to 
\begin{align}\label{Null Infinity: normality conditions2'}
	(n-3) \xi_{AB}\big|_{tf} &= F_{\go}{}^C{}_{ACB}\big|_{tf}, & (n-2) \xi^C{}_C = -\tfrac{1}{2} F_{\go}{}^{CD}{}_{CD},
\end{align}
and
\begin{align}\label{Null Infinity: normality conditions3'}
	(n-2)\psi_A &= -\tfrac{1}{2} \covD^B C_{BA}.
\end{align}
In dimension $n=3$ the first equation in \eqref{Null Infinity: normality conditions2'} is identically satisfied, the second determines $\xi^C{}_C$ and \eqref{Null Infinity: normality conditions3'} fixes $\psi_A$. The only component of the connection remaining unconstrained is therefore the trace-free symmetric part of $C_{AB}$. A direct computation then shows that the transformation rules \eqref{Null Infinity: gauge transformation of fields - C} for $C_{(AB)}$ coincide with \eqref{Null infinity: C transformation rules}. In other terms, it defines a Poincaré operator \eqref{Null infinity: Poincaré operator}.

We therefore obtained the following Proposition:
\begin{Proposition}{\cite{herfray_asymptotic_2020}}\label{Null Infinity: proposition, Poincaré connection}
	Let $\left(\scrI \to \gS, [n^{\mu}, h_{\mu\nu}]\right)$ be a $3$-dimensional conformal Carrollian geometry, there is a one-to-one correspondence between, on the one hand, $\iso\left(3,1\right)$-valued connections which are compatible with the conformal Carroll geometry, compatible with Thomas operator, torsion-free and normal and, on the other hand, choice of Poincaré operators $\cP$ as defined in section \ref{sss: Strongly conformally Carrollian geometry}.
	
	In other terms strongly conformally Carrollian geometry  $\left(\scrI \to \gS, [n^{\mu}, h_{\mu\nu}],\cP\right)$ in dimension $3$ are in one-to-one correspondence with such $\iso\left(3,1\right)$-valued connections.
\end{Proposition}

In dimension $n\geq4$ the situation is similar but \eqref{Null Infinity: normality conditions1'} and \eqref{Null Infinity: normality conditions2'} together constrain the ``news tensor'' $\LieD_{n} C_{AB}$ and therefore constrain the admissible Poincaré operator.

In dimension $n=2$, the situation is radically different. Since one-dimensional trace-free symmetric tensors vanish identically there is no freedom in $C_{AB}$, however equation \eqref{Null Infinity: normality conditions2'} and \eqref{Null Infinity: normality conditions3'} are now identically satisfied and do not constrain $\xi^C{}_C$ and $\psi_{A}$.  Thus $\xi^C{}_C$ and $\psi_A$ are the free parameters for the connection. In fact one can show \cite{herfray_asymptotic_2020,herfray_tractor_2022} that $M := \tfrac{2}{n-1} \xi^C{}_C$ and $N_A := -\psi_A$ then respectively correspond to choices of mass and angular momentum aspects for the corresponding three-dimensional asymptotically flat spacetime.

Finally, from the fundamental theorem of Cartan geometry \eqref{Introduction: fundamental theorem}, we have:
\begin{Proposition}
	Let $\left(\scrI \to \gS, [n^{\mu}, h_{\mu\nu}], \cP, \right)$ be a $n$-dimensional strongly conformally Carrollian geometry, it is locally isomorphic to the model \eqref{Null infinity: homogeneous space realisation} if and only if the curvature \eqref{Null infinity: curvature} of the corresponding $\iso\left(n,1\right)$-valued connection vanishes.
	
	 This then implies that the algebra of local symmetries is $\iso\left(n,1\right)$.
\end{Proposition}

\subsubsection{Poincaré operator revisited}

Let $\left(\scrI \to \gS, [n^{\mu}, h_{\mu\nu}], \cP\right)$ be a strongly conformally Carrollian geometry and let $D$ be the corresponding $\iso\left(n,1\right)$-valued connection. Its action on dual tractors is
\begin{equation*}
D\Phi_I \xeq{e} \Mtx{
	d +\ga & -\gth^B  & 0 & -l \\
	-\xi_A& d^{\go} &  \gth_A & \tfrac{1}{2}C_A \\
	0 & \xi^B & d -\ga & \psi \\
	0 & 0 &  0 &d 
}\Mtx{\Phi_- \\ \Phi_B \\ \Phi_+ \\ \Phi_0}
\end{equation*} 
where duality is defined by $\Phi_I \Phi^I = \Phi_+\Phi^+ + \Phi_A\Phi^A + \Phi_-\Phi^- + \Phi_0\Phi^0$.

Let $\Phi_I = \left(f , \Phi_A, \Phi_+, \Phi_0\right)$ be a covariantly constant dual tractor $D_{\gr}\Phi_I=0$. Making use of \eqref{Null infinity: compatibility with Thomas operator}, the equation $D\Phi_+ =0$ is found to be equivalent to $\Phi_A = \covD_A f$, $\Phi_0 = \covD_0^{(\gTh)}f$. Then
\begin{align}\label{Null Infinity: Poincaré operator from tractors}
&D_{0}\Phi_0 = \covD_{0}\covD_0^{(\gTh)}f \nonumber\\
&D_{0}\Phi_A = D_A \Phi_0 = \covD_{A}\covD_0^{(\gTh)}f \\
&D_{A}\Phi_B = \covD_{A}\covD_{B}f +\tfrac{1}{2}C_{BA} \fd -\left(\xi_{AB} + \tfrac{\gTh}{4} C_{AB}\right) f + h_{AB} \Phi_+. \nonumber
\end{align}
Making use of the normality condition \eqref{Null Infinity: normality conditions1'} we find that \eqref{Null Infinity: Poincaré operator from tractors} coincides with the Poincaré operator \eqref{Null infinity: Poincaré operator} (in particular one can check, using the identity $\covD_{[A}\covD_{B]}f = -\tfrac{1}{2}\fd (dl)_{AB}  $ and torsion-freeness, that $D_{(A}\Phi_{B)} = D_{A}\Phi_{B}$). This can be taken as a direct proof that the Poincaré operator is well-defined or as an alternative definition.

We can now give a geometrical meaning to zeros of the Poincaré operator. Solving for $\Phi_+$ with $D^{C}\Phi_C=0$ and making use of the normality condition \eqref{Null Infinity: normality conditions2'} we obtain that $\Phi_I$ is the image of $f$ by Thomas operator \eqref{Null Infinity: Thomas operator} : $\Phi_I = T\left(f\right)_I$. It follows from the previous discussion that if $T\left(f\right)_I$ is covariantly constant with respect to a normal tractor connection then $f$ must be a zero of the corresponding Poincaré operator. In fact one can prove \cite{herfray_asymptotic_2020} that $f$ is a zero of the Poincaré operator if and only if $T\left(f\right)_I$ is covariantly constant (this amounts to proving that $D_{\gr}\Phi_+$ identically vanishes when the other components of $D\Phi_I$ are zero).

\subsubsection{Space-time interpretation}

As was previously highlighted the null-tractor bundle of the conformal boundary $\scrI = \parD M$ of an asymptotically flat spacetime can be canonically identified with the restriction of the usual tractor bundle of $M$.

The (unique) normal tractor connection of an asymptotically flat spacetimes then induces a connection on the null-tractor bundle and, assuming standard fall-off on the energy-momentum tensor of $M$, one can prove that this connection is normal \cite{herfray_tractor_2022}. Therefore the boundary of an asymptotically flat-spacetime is always equipped with a normal Cartan connection modelled on \eqref{Null infinity: homogeneous space realisation} or, equivalently, with a strongly conformally Carrollian geometry $\left(\scrI\to \gS, [n^{\mu},h_{\mu\nu}],\cP\right)$. Once again, this essentially boils down to the fact that, in dimension $d=4$, the asymptotic shear of an asymptotically flat spacetime as the correct transformation rules \eqref{Null infinity: C transformation rules} to define a Poincaré operator. Note however that this results extends to any dimension $d\geq3$, in particular it unifies the role of the asymptotic shear in $d\geq4$ with that of the mass ans angular momentum aspects in $d=3$.

Let us now discuss asymptotic symmetries. First note that while the BMS group is the group of symmetries of a (weak) conformal Carrollian geometry, the Poincaré group is the group of symmetries of a flat strongly conformal Carrollian geometry - This follows from the fundamental theorem of Cartan geometry \ref{Introduction: fundamental theorem}. Flatness is essential here, a generic geometry will have no symmetries. Now if $\left(\scrI\to \gS, [n^{\mu},h_{\mu\nu}],\cP\right)$ is the strongly conformal Carrollian geometry induced at the boundary of an asymptotically flat spacetime, one can prove \cite{herfray_tractor_2022} that the curvature of the induced connection correspond in spacetime dimension $d=4$ to the Newman--Penrose coefficients $\Psi^0_4$, $\Psi^0_3$, $Im\left(\Psi^0_2\right)$. In other terms, as was highlighted in the introduction, gravitational radiation can be geometrically understood as the curvature of the induced Cartan connections i.e. gravitational waves are the obstruction to being able to identify null infinity with the model \eqref{Null infinity: homogeneous space realisation}. These identifications are equivalent to choosing an action of the Poincaré group on $\scrI$, and therefore ``gravitational radiation is the obstruction to having a preferred Poincaré group inside the BMS group'' in the precise sense of Cartan geometry.

We stress that this interplay between gravitational radiation and the induced Cartan geometry is a specific feature of the critical dimension $d=4$. In dimension $d=3$, Einstein equations force the curvature of the induced Cartan geometry to vanish (the corresponding equations on the mass and angular momentum aspects are the so-called ``conservation equation'', see \cite{herfray_tractor_2022}). This is in line with the absence of gravitational radiation for $d=3$. In dimension $d\geq5$, Einstein equations again imply (assuming enough differentiability) that the curvature of the induced Cartan geometry vanishes. This follows from results of \cite{hollands_bms_2017} and corresponds to the fact that radiative degrees of freedom in these higher dimensions are not found in the asymptotic shear but rather sit at lower order in the asymptotic expansion and therefore cannot be captured by the boundary tractor connection. 

\subsection{Symmetries, BMS coordinates and transformation properties of the shear}

By construction, diffeomorphisms of $\scrI$ and $\bbR \times \ISO\left(n-1\right)$-gauge transformations all naturally act on the space of tractor connections given by Proposition \eqref{Null Infinity: proposition, Poincaré connection}, (this would form the symmetry group for the corresponding Chern-Simons action at null-infinity, see \cite{nguyen_effective_2021}). If one considers a fixed conformal Carrollian geometry $\left(\scrI\to \gS , [n^{\mu}, h_{\mu\nu}]\right)$ symmetries consist of the BMS group \eqref{Null infinity: BMS symmetries} supplemented by $\bbR \times \ISO\left(n-1\right)$-gauge transformations. These gauge transformations correspond to change of tetrads \eqref{Null infinity: tetrad transformation rules} and, in particular the $\bbR$-factor corresponds to Weyl rescalings: one can therefore think of this as an extension of the Weyl-BMS group\footnote{With our definition for conformal Carrollian geometry diffeomorphisms of the sphere are not symmetries, one can however obtain the generalised BMS group from \cite{campiglia_asymptotic_2014} by relaxing the definition and only requiring $\left(\scrI\to \gS , n^{\mu}\right)$ with the equivalence class replaced by a weighted vector field $n^{\mu} \in \So{T\scrI \otimes L^{-1}}$.} of \cite{barnich_aspects_2010,barnich_finite_2016, freidel_weyl_2021}. The action of this group on the asymptotic shear is for example obtained by composing diffeomorphisms of $\scrI$ with the gauge transformations \eqref{Null Infinity: gauge transformation of fields - C}, \eqref{Null Infinity: tractor transformation rules r_A t_+}.

We now discuss how this group can be reduced to the more usual symmetry groups by choice of BMS coordinates. Let $ u \from \scrI \to \bbR$ be a trivialisation of $\scrI \to \bbR$ let us choose a tetrad $\left\{e_A{}^{\mu}, n^{\mu}\right\}$ such that $e_A^{\mu}(du)_{\mu}=0$. Changing the trivialisation $u \mapsto \uh$ gives
\begin{align*}
\begin{array}{lcl}
e_A{}^{\mu} &\mapsto& \gl^{-1} \; m_A{}^B\left( \;e_B{}^{\mu}-  t_B n^{\mu}\right)\\
n^{\mu} & \mapsto& \gl^{-1}\; n^{\mu}
\end{array}&&\quad t_B= \gl^{-1}\covD_B \uh.
\end{align*}
This procedure thus reduces $\bbR \times \ISO\left(n-1\right)$-gauge transformations to $\bbR \times \SO\left(n-1\right)$, with the remaining freedom corresponding to Weyl rescalings and rotations of the tetrad in the direction transverse to $n^{\mu}$. We can fix the Weyl rescalings as follows: Since $u$ is a trivialisation, $n^{\mu} (du)_{\mu}$ is nowhere vanishing and there is a unique representative $n^{\mu} \in [n^{\mu}]$ such that $n^{\mu}(du)_{\mu}=1$. To preserve this gauge fixing, the above transformation rules must then be complemented by
\begin{equation*}
\gl =  \LieD_{n} \uh.
\end{equation*}
This last step reduces the gauge freedom to $\SO\left(n-1\right)$.

In these type of gauges, the dual-tetrad element $l_{\mu} = (du)_{\mu}$ satisfies $dl=0$, $\LieD_{n}l=0$ which drastically simplifies the expressions for torsion-freeness and normality of the tractor connection. In particular we then have $C_B = C_{AB} \gth^B$, where $C_{AB}$ is trace-free symmetric and follows the transformation rules
\begin{align*}
\tfrac{1}{2}C_{AB} \quad \mapsto \quad   \gl^{-1}m_A{}^C m_B{}^D  \left( \tfrac{1}{2}C_{CD} +  \gl^{-1}\covD_C\covD_D \uh -2\gl^{-1} \gU_{(C} \covD_{D)} \uh + \gl^{-2} \left( \tfrac{\gld}{\gl} + \tfrac{\gTh}{2} \right) \covD_C \uh \covD_D\uh \right)\Big|_{tf}
\end{align*}
where we made use of \eqref{Null Infinity: gauge transformation of fields - C}, \eqref{Null Infinity: tractor transformation rules r_A t_+} and $t_B= \gl^{-1}\covD_B \uh$. These are the finite (i.e. non infinitesimal) transformations property of the asymptotic shear under the Weyl-BMS group \cite{barnich_finite_2016, freidel_weyl_2021,herfray_tractor_2022}.

Let us anticipate on the next section and suppose that we choose a Bondi gauge i.e. a representative $h_{\mu\nu} \in  [h_{\mu\nu}]$ such that $\LieD_{n}h_{\mu\nu}=0$. Then $\gTh = 0$ and Weyl rescalings preserving this choice must satisfy $\gld=0$. What is more if $u$ and $\uh$ are two trivialisations satisfying $n^{\mu}(du)_{\mu} = n^{\mu} (d\uh)_{\mu}=0$ we must have $\uh = \gl u + \pi^*f$ with $f$ is a function on $\gS$. This reduces the above transformation rules accordingly.
	
\section{Null Infinity in Bondi gauge}

Even though the geometry of null infinity is based on a \emph{conformal} Carrollian geometry it is customary in the literature (see e.g. the reviews \cite{ashtekar_geometry_2015,madler_bondi-sachs_2016,ashtekar_null_2018}) to work in a fixed gauge $h_{\mu\nu}\in [h_{\mu\nu}]$ satisfying\footnote{Choosing this gauge is convenient but of course not necessary, see e.g. \cite{barnich_aspects_2010,barnich_finite_2016,frauendiener_new_2021} for detailed discussions not assuming this gauge.} $\LieD_{n}h_{\mu\nu}=0$. We will call these representatives ``Bondi gauge'' even though, strictly speaking, in the literature Bondi gauge corresponds to the situation where $h_{\mu\nu}$ has constant curvature. This will however be clear from the discussion when we need this extra requirement or not. 

With our definitions \ref{Carrollian spacetimes: Carrolian geometry def} and \ref{Null infinity: Conformal Carroll geometry def} a conformally Carrollian geometry in Bondi gauge is in fact simply Carrollian. One might therefore be under the impression that, working in Bondi gauge, null infinity is just Carrollian geometry. This is however misleading because the \emph{strong} geometries do not match, even in Bondi gauge \cite{ashtekar_radiative_1981,ashtekar_geometry_2015}. It follows from propositions \ref{Carrolian spacetimes: proposition, Carrolian connection} and \ref{Null Infinity: proposition, Poincaré connection} that the corresponding Cartan geometries cannot be the same either.
 
 The deep reason for the discrepancy can be traced back to the existence of $n+1$ translations on the homogeneous realisation of null infinity \eqref{Null infinity: homogeneous space realisation}: these $n+1$ translations must still be present in Bondi gauge and do not match the $n$ Carrollian translations of the homogeneous models \eqref{Carrollian spacetimes: homogeneous space realisation}. We will here discuss the implications of this fact for the underlying Cartan geometries (the main results are summarised in Table \ref{Table: Bondi gauge}). 


\begin{table}
	\caption{Cartan geometry of null infinity in Bondi gauge}\label{Table: Bondi gauge}
\begin{center}
\begin{tabular}{p{4cm}cc}
	\hline\\
	\multicolumn{1}{l}{\textbf{Model}} &  &\\[0.5em]
	\multicolumn{3}{c}{$\scrI_{B}^{(n)}=$ $\qquad \Quotient{\bbR \times \Carr_{\gL}\left(n\right)}{\bbR \times  \ISO\left(n-1\right)}$} \\ [1.5em]
	
	\hline\\
	\multicolumn{3}{l}{\textbf{(Weak) conformally Carrollian geometry in Bondi gauge}} \\[1.5em]
	$\left(\scrI \to \gS, n^{\mu}, h_{\mu\nu} \right)$ & defines a canonical & $\bbR \times \ISO\left(n-1\right)$-principal bundle \\[1em]
	& has symmetry group & $\Co{\gS} \rtimes \ISO\left(\gS, h_{\mu\nu}\right)$ \\[1em] \hline\\
	\multicolumn{3}{l}{\textbf{Strong conformally Carrollian geometry in Bondi gauge}}\\[1.5em]
	$\left(\scrI \to \gS, n^{\mu}, h_{\mu\nu}, [\covD_{\gr}] \right)$& defines a unique & $\iso\left(n,1\right)$-valued connection $D$ \\ \\
	\begin{tabular}{c}
		Flatness of $D$ \\
		\underline{and} \\ Constant curvature $h_{\mu\nu}$\end{tabular} &  \begin{tabular}{c}
		reduces the \\symmetry group to
	\end{tabular}  & $\bbR \times \Carr_{\gL}\left(n\right)$. \\[2em]
\hline
\end{tabular}
\end{center}
\end{table}

\subsection{Null Infinity in Bondi gauge: homogeneous model}

The base manifold of the homogeneous model \eqref{Null infinity: homogeneous space realisation} is the conformal sphere $S^{n-1}$. It can be thought as the conformal compactification of the constant curvature Riemannian models : the round sphere $S^{n-1}$, the euclidean plane $\bbR^{n-1}$ and hyperbolic space $H^{n-1}$. From the homogeneous perspective, this simply corresponds to the fact that their isometry groups \eqref{Carrollian spacetimes: Riemannian Isometry Groups}, which we collectively write $\ISO_{\gL}\left(n-1\right)$, are all subgroups of the conformal group $\SO\left(n,1\right)_0$. Restricting the $\SO\left(n,1\right)_0$ part of $\ISO\left(n,1\right) = \bbR^{n+1} \rtimes \SO\left(n,1\right)_0$ to these isometry groups, one obtains the following model spaces:
\begin{align}\label{Null Infinity in Bondi gauge: Homogeneous model}
\scrI_{Bondi_{\gL}}^{(n)} = \Quotient{\bbR^{n+1}\rtimes \ISO_{\gL}\left(n-1\right)}{\bbR \rtimes \ISO\left(n-1\right)}.
\end{align}
For $\gL>0$ we obtain the homogeneous model for null infinity in Bondi gauge $\scrI_{Bondi_{\gL>0}}{}^{(n)} \simeq \bbR \times S^{n-1}$. The homogeneous space \eqref{Null Infinity in Bondi gauge: Homogeneous model} can be rewritten in a form which makes the relation to the Carrollian homogeneous model \eqref{Carrollian spacetimes: homogeneous space realisation} stand out
\begin{align}\label{Null Infinity in Bondi gauge: Homogeneous model2}
	\scrI_{Bondi_{\gL}}^{(n)} = \Quotient{\bbR \times \Carr_{\gL}\left(n\right)}{\bbR \times \ISO\left(n-1\right)}.
\end{align}
The geometry of null infinity in Bondi gauge is therefore very close that of a Carrollian manifold, it however enjoys an extra abelian symmetry. At the level of Lie algebras,
{\small 
	\begin{align}\label{Null infinty in Bondi gauge: Lie algebras1}
	 &\bbR \oplus \carr_{\gL}\left(n\right) = \left \{\quad \begin{pmatrix} 0 & -m_B & 0 &0
	\\ \gL m^A & \fkm^A{}_B &  m^A &0
	\\	0 & -\gL m_B & 0&0
	\\ t_+ & t_B &  t_- &0 \end{pmatrix}  \quad \Big|\quad  \begin{array}{c} \fkm^A{}_{B} \in \so(n-1),\;  t_B \in \bbR^{n-1},  \\[0.4em]   (m^A, t_-) \in \bbR^{n+1}, \, t_+ \in \bbR 
	\end{array}  \quad    \right \},
	\\\nonumber\\
&\bbR \oplus \iso\left(n-1\right)  = \left \{\quad
	\begin{pmatrix} 0 & 0 & 0 &0
	\\0 & \fkm^A{}_B &  0 &0
	\\	0 & 0 & 0&0
	\\ t_+ & t_B &  0 &0  \end{pmatrix} \quad \Big|\quad  \begin{array}{c} \fkm^A{}_{B} \in \so(n-1),  \\[0.4em] \;  t_B \in \bbR^{n-1}, \; t_+ \in \bbR 
	\end{array}  \quad    \right \},
	\end{align}}{}\\
 the extra symmetry corresponds to the presence of an extra  $t_+$ term as compare to the Carrollian case \eqref{Carrollian spacetimes: Carroll Lie algebra}, \eqref{Carrollian spacetimes: Carroll stabilisator Group}. As compare to null infinity \eqref{Null infinty: Poincaré Lie algebra}, \eqref{Null infinity: Poincaré stabilisator Group} this corresponds to a remaining Carrollian special conformal transformation.

We stress that the abelian factor in $\bbR \times \Carr_{\gL}\left(n\right)$ does act non-trivially on the homogeneous space (despite the fact that the, misleading, expression \eqref{Null Infinity in Bondi gauge: Homogeneous model2} might suggest the opposite). This can be realised explicitly by considering
{\small \begin{equation}\label{Null infinity in Bondi gauge: ambient space realisation}
	\scrI^{(n)}_{Bondi_{\gL}} = \left\{Y^I = \Mtx{Y^+ \\ Y^A \\ Y^- \\ Y^0} \in \bbR^{n+2} \quad\big|\quad Y^I Y^J h_{IJ} =0,\quad  Y^I I^J_{(\gL)} h_{IJ} = 1   \right\}
	\end{equation}}{}\\
where $I^I_{(\gL)} =\left(1 , 0^A, -\gL, 0\right)$ explicitly breaks conformal invariance as compare to \eqref{Null infinity: ambient space realisation}. The transitive action of $\bbR \times \Carr_{\gL}\left(n\right)$ on \eqref{Null infinity in Bondi gauge: ambient space realisation} is then realised by restricting the action of the Poincaré group on \eqref{Null infinity: ambient space realisation} to the subgroup stabilising $I_{(\gL)}$. Note that even though \eqref{Carrollian spacetimes: ambient space realisation} and \eqref{Null infinity in Bondi gauge: ambient space realisation} coincide as manifolds the group action is different and they therefore differ as homogeneous spaces.

\begin{Extra}
	\clearpage
	
\begin{tcolorbox}
{\small

Let us write $\bbR^{n+2}$ as $Y = \left(Y^{t}, Y^A, Y^{n}, Y^0\right)$ with degenerate metric
\begin{equation*}
Y^2 = -(Y^t)^2 + Y^A Y_A + (Y^{n})^2
\end{equation*}
let us pick the time-like vector $I = \left( 1 , 0, 0,0\right)$, we have	
{\small \begin{equation*}
	\scrI^{(n)}_{B} = \left\{Y = \Mtx{Y^t \\ Y^A \\ Y^{n} \\ Y^0} \in \bbR^{n+2} \quad\big|\quad Y^2 =0,\quad  Y.I = 1   \right\}
	\end{equation*}}
i.e.
{\small \begin{equation*}
\scrI^{(n)}_{B} = \left\{Y = \Mtx{1 \\ Y^A \\ Y^{n} \\ Y^0} \in \bbR^{n+2} \quad\big|\quad (Y^A)^2 + (Y^n)^2 =1  \right\} = \bbR \times S^{n-1}
\end{equation*}}
The corresponding isometry group and stabiliser of $X = \left( 1 , 0, 1,0\right)$ are
\begin{align*}
&\begin{pmatrix}
1 & 0 & 0 & 0\\
0 & m^A{}_B & m^A{}_n &0\\
0 & m^n{}_B & m^n{}_n & 0\\
t_0 & t_A & t_n & 1
\end{pmatrix}  & &\begin{pmatrix}
1 & 0 & 0 & 0\\
0 & \mb^A{}_B & 0 &0\\
0 & 0 & 1 & 0\\
\tb_0 & \tb_A & -\tb_0 & 1
\end{pmatrix}
\end{align*}
Note that even though the isomorphisms
\begin{align*}
&\begin{pmatrix}
1 & 0 & 0 & 0\\
0 & m^A{}_B & m^A{}_n &0\\
0 & m^n{}_B & m^n{}_n & 0\\
t_0 & t_A & t_n & 1
\end{pmatrix}  &\mapsto & 
&\begin{pmatrix}
1 & 0 & 0 & 0\\
0 & m^A{}_B & m^A{}_n &0\\
0 & m^n{}_B & m^n{}_n & 0\\
0 & t_A & t_n & 1
\end{pmatrix} \\ 
&\begin{pmatrix}
1 & 0 & 0 & 0\\
0 & \mb^A{}_B & 0 &0\\
0 & 0 & 1 & 0\\
\tb_0 & \tb_A & -\tb_0 & 1
\end{pmatrix} 
&\mapsto & 
&\begin{pmatrix}
1 & 0 & 0 & 0\\
0 & \mb^A{}_B & 0 &0\\
0 & 0 & 1 & 0\\
0 & \tb_A & 0 & 1
\end{pmatrix}
\end{align*}
are canonical, at the level of the quotient

\begin{align*}
& \begin{pmatrix}
1 & 0 & 0 & 0\\
0 & \gd^A{}_B & 0 &0\\
0 & 0 & 1 & 0\\
t_0 +\tb_0  & t_A + \tb_A & t_n -\tb_0 & 1
\end{pmatrix} &\mapsto &&  \begin{pmatrix}
1 & 0 & 0 & 0\\
0& \gd^A{}_B & 0 &0\\
0& 0 & 1 & 0\\
0&t_A + \tb_A & t_n- t_0 & 1
\end{pmatrix} \neq  \begin{pmatrix}
1 & 0 & 0 & 0\\
0& \gd^A{}_B & 0 &0\\
0& 0 & 1 & 0\\
0&t_A + \tb_A & t_n& 1
\end{pmatrix}
\end{align*}
this is not the case.

It is natural to choose representatives of the form
\begin{align*}
	& \begin{pmatrix}
		1 & 0 & 0 & 0\\
		0 & \gd^A{}_B & 0 &0\\
		0 & 0 & 1 & 0\\
		0 & 0 & u & 1
	\end{pmatrix}
\end{align*}
then the abelian factor does act non trivially:
\begin{equation*}
	\begin{pmatrix}
		1 & 0 & 0 & 0\\
		0 & \gd^A{}_B & 0 &0\\
		0 & 0 & 1 & 0\\
		\ga & 0 & 0 & 1
	\end{pmatrix}
\begin{pmatrix}
	1 & 0 & 0 & 0\\
	0 & \gd^A{}_B & 0 &0\\
	0 & 0 & 1 & 0\\
	0 & 0 & u & 1
\end{pmatrix} =  \begin{pmatrix}
1 & 0 & 0 & 0\\
0 & \gd^A{}_B & 0 &0\\
0 & 0 & 1 & 0\\
\ga & 0 & u & 1
\end{pmatrix} \sim \begin{pmatrix}
1 & 0 & 0 & 0\\
0 & \gd^A{}_B & 0 &0\\
0 & 0 & 1 & 0\\
0 & 0 & u + \ga & 1
\end{pmatrix}
\end{equation*}

}
\end{tcolorbox}
\end{Extra}

\subsection{Null Infinity in Bondi gauge: the principal bundle}

Let $\left(\scrI \to \gS, [n^{\mu}, h_{\mu\nu}] \right)$ be a conformal Carrollian geometry. As discussed in section \ref{ss: The null-tractor bundle} it is canonically equipped with a $\bbR^{n} \rtimes \left(\bbR^* \times \ISO\left(n-1\right)\right)$-principal bundle - the bundle of orthonormal tractor frames. What is more, any choice of tetrad $\{e_A{}^{\mu},n^{\mu}\}$ defines a splitting isomorphism $\cT$ with transformation rules given by \eqref{Null Infinity: tractor transformation rules} and \eqref{Null Infinity: tractor transformation rules r_A t_+}. If we work in Bondi gauge, i.e. choose a metric representative $h_{\mu\nu}$ such that $\LieD_{n}h_{\mu\nu}=0$ then corresponding tetrads are unique up to
\begin{align}\label{Null infinity in Bondi gauge: tetrad transformation rules}
\begin{array}{lcl}
e_A{}^{\mu} &\mapsto& m_A{}^B\left( \;e_B{}^{\mu}-  t_B n^{\mu}\right)\\
n^{\mu} & \mapsto& n^{\mu}
\end{array}&& \Mtx{m^A{}_B & 0 \\ t_B &1}&\in \ISO\left(n-1\right),
\end{align}
and the tractor transformation rules \eqref{Null Infinity: tractor transformation rules} reduce to
\begin{equation}\label{Null Infinity in Bondi gauge: tractor transformation rules}
\Mtx{\Phi^+ \\ \Phi^A \\ \Phi^- \\ \Phi^0} \mapsto \Mtx{1 & 0 & 0& 0\\ 0 & m^A{}_B & 0 & 0 \\ 0 & 0& 1& 0 \\ t_+&t_B&0&1} \Mtx{\Phi^+ \\ \Phi^B \\ \Phi^- \\ \Phi^0}
\end{equation}
with
\begin{align}\label{Null Infinity in Bondi gauge: tractor transformation rules t_+}
t_+ &= \tfrac{1}{n-1}\left( -\covD_C t^C + \tfrac{1}{2}\LieD_{n}(t^2) + t^C (\LieD_{n}l)_C\right).
\end{align}
The appearance of a non-zero $t_+$ term in the above transformation rules means that the $\bbR^{n} \rtimes \left(\bbR \times \ISO\left(n-1\right)\right)$-principal bundle of null infinity reduces to an $\bbR \times \ISO\left(n-1\right)$-principal bundle. In particular the resulting principal bundle does \emph{not} coincide with the obvious $\ISO\left(n-1\right)$-principal bundle \eqref{Null infinity in Bondi gauge: tetrad transformation rules} of Carrollian geometry.

In consequence, there are always \emph{two} principal bundles which can be constructed from a Carrollian geometry, the first is obtained as a reduction from the frame bundle and has structure group $\ISO\left(n-1\right)$, the second is obtained from frames for 1-jets of functions (the construction is similar to the one described in section \ref{sss: Null-tractors: invariant definition}) and has structure group $\bbR \times \ISO\left(n-1\right)$.

\subsection{From Poincaré operator to equivalence class of connections}

Let $\left( \scrI \to \gS, [n^{\mu}, h_{\mu\nu}], \cP\right)$ be a strongly conformally Carrollian geometry. Let $h_{\mu\nu} \in [h_{\mu\nu}]$ be a Bondi gauge (in particular $\gTh := \tfrac{1}{n-1}h^{\mu\nu}\hd_{\mu\nu}=0$), the Poincaré operator then correspond to a choice of trace-free symmetric tensor $C_{AB}$ following the transformation rules \eqref{Null infinity: C transformation rules} with the restriction $\gl=1$ to preserve the Bondi gauge.

The transformation rules obtained in this way are the transformation rules for the components \eqref{Carrollian spacetimes: affine connection coordinates} of a torsion-free connection $\covD_{\gr}$ compatible with the Carrollian geometry $\left(n^{\mu} , h_{\mu\nu} \right)$. However since a Poincaré operator only is equivalent to a \emph{trace-free} tensor we are missing a trace-part to uniquely define this connection. Rather a choice of Poincaré operator in a fixed Bondi gauge corresponds to an equivalence class $[\covD_{\gr}]$ of connections
\begin{align*}
\covD_{\gr} &\sim \covDt_{\gr} &&\Leftrightarrow& \covD_{\gr}-\covDt_{\gr} &\propto h_{\gr\nu}n^{\mu}.
\end{align*}
In a tetrad this equivalence relation indeed amounts to shifting the trace of $C_{AB}$.

Such equivalence classes were called radiative structure in \cite{ashtekar_geometry_2015} and have been identified as the initial characteristic data for gravity at null infinity in \cite{geroch_asymptotic_1977,ashtekar_radiative_1981}. 

From this perspective, the discrepancy between the flat model for null infinity in Bondi gauge \eqref{Null Infinity in Bondi gauge: Homogeneous model2} and the homogeneous Carroll space-times \eqref{Carrollian spacetimes: homogeneous space realisation} receives a clear geometrical interpretation: while flat Carrollian geometries are equipped with a preferred torsion-free connection $\covD_{\gr}$, the flat model for null infinity in Bondi gauge only posses an equivalence class $[\covD_{\gr}]$ of such objects  and the extra abelian symmetry corresponds to a shift from one connection representative to another.

\subsection{Null Infinity in Bondi gauge: the connection}

Let $\left(\scrI \to \gS, [n^{\mu}, h_{\mu\nu}], \cP \right)$ be a strongly conformal Carrollian geometry. By results of the previous section this is equivalent to a choice of Cartan geometry modelled on \eqref{Null infinity: homogeneous space realisation}. Let $h_{\mu\nu}\in [ h_{\mu\nu}]$ be a Bondi gauge and let $\{e_A{}^{\mu},n^{\mu}\}$ be a choice of adapted tetrad. The corresponding Cartan connection is of the form
\begin{equation}\label{Null infinity in Bondi gauge: the connection}
	D\Phi^I \xeq{e} \Mtx{
		d +\ga & -\gth_B  & 0 & 0 \\
		-\xi_A& d^{\go} &  \gth^A &0 \\
		0 & \xi_B & d -\ga & 0 \\
		-\psi & -\tfrac{1}{2}C_B &  l &d 
	}\Mtx{\Phi^+ \\ \Phi^A \\ \Phi^- \\ \Phi^0}.
\end{equation} 
and the action of the $\bbR \times \ISO\left(n-1\right)$-valued gauge transformations \eqref{Null Infinity in Bondi gauge: tractor transformation rules} is given by
\begin{align}\label{Null Infinity in Bondi gauge: gauge transformation of fields}
	\gth^A &\mapsto m^A{}_B\gth^B,  & l &\mapsto l + \gth^C t_C, & \go^A{}_B &\mapsto m^A{}_C \go^C{}_D m_B{}^D - dm^A{}_C m_B{}^C,\nonumber\\ \nonumber\\
	\xi^A &\mapsto m^A{}_B\xi^B,& \ga &\mapsto \ga,& -\tfrac{1}{2}C_A & \mapsto m_A{}^B\left( -\tfrac{1}{2}C_B -d^{\go}t_B - \gth_B t_+\right),
\end{align}
\begin{align*}
	\psi& \mapsto \psi +\xi^C t_C + d_{-\ga}t_+.
\end{align*}
Supposing that the connection is torsion-free, the ``compatibility with Thomas operator'' conditions \eqref{Null infinity in Bondi gauge: compatibility with Thomas operator} now are
\begin{align}\label{Null infinity in Bondi gauge: compatibility with Thomas operator}
	\ga &=0, & 
	\tfrac{1}{2} h^{CD} C_{CD} & =0.
\end{align}
Here again the constraints \eqref{Null Infinity in Bondi gauge: tractor transformation rules t_+} on the transformation rules can be understood as the subspace of gauge transformations preserving the gauge fixing condition \eqref{Null infinity in Bondi gauge: compatibility with Thomas operator}).

The appearance of the extra $t_+$ gauge parameter in the gauge transformation \eqref{Null Infinity in Bondi gauge: gauge transformation of fields} as compare to those of a Carrollian-valued connection \eqref{Carrollian spacetimes: gauge transformation of fields - C} implies an essential difference with the strongly Carrollian case: the pair $\left(\go^A{}_B, C_B\right)$ does not define a unique connection $\covD$ on the tangent bundle $T\scrI$, rather we have an equivalence class of connections $[\covD]$ where representatives are related by gauge transformations
\begin{equation*}
	\covD_{\gr} \mapsto \covD_{\gr} - (t_+) h_{\gr\nu} n^{\mu}.
\end{equation*}
This is in line with the preceding discussion on Poincaré operators.

A choice of equivalence class $\left(\scrI \to \gS , n^{\mu}, h_{\mu\nu}, [\covD]  \right)$ always defines a strongly conformally Carrollian geometry and therefore always defines a Cartan geometry modelled on \eqref{Null infinity: homogeneous space realisation}. However there does not seem to have any Cartan geometry modelled on \eqref{Null Infinity in Bondi gauge: Homogeneous model} generically equivalent to such gauged-fixed geometry $\left(\scrI \to \gS , n^{\mu}, h_{\mu\nu}, [\covD]  \right)$. Nevertheless, if the Cartan connection \eqref{Null infinity in Bondi gauge: the connection} is flat and the representative $h_{\mu\nu}$ satisfies the constant curvature condition 
\begin{equation*}
F_{\go}^A{}_B -2\gL \gth^A \W \gth_B=0,
\end{equation*}
then it follows from the normality conditions \eqref{Null Infinity: normality conditions2'} that the connection in fact takes values in the Lie algebra \eqref{Null infinty in Bondi gauge: Lie algebras1} of $\bbR \times \Carr_{\gL}\left(n\right)$:  \begin{equation*}
D\Phi^I = \Mtx{
	d & -\gth_B  & 0 & 0 \\
	\gL \gth^A & d^{\go} &  \gth^A &0 \\
	0 & -\gL \gth_B & d & 0 \\
	-\psi & -\tfrac{1}{2}C_B &  l &d 
}\Mtx{\Phi^+ \\ \Phi^A \\ \Phi^- \\ \Phi^0}.
\end{equation*} 
From the fundamental theorem of Cartan geometry we therefore have:
\begin{Proposition}
	Let $\left(\scrI \to \gS, n^{\mu}, h_{\mu\nu}, [\covD] \right)$ be a strongly conformally Carrolian geometry such that the metric representative satisfies \begin{equation*}
	F_{\go}^A{}_B -2\gL \gth^A \W \gth_B=0
	\end{equation*} and such that the corresponding Cartan connection is flat. Then $\left(\scrI \to \gS, n^{\mu}, h_{\mu\nu}, [\covD]\right)$ is locally isomorphic to the homogeneous space model \eqref{Null Infinity in Bondi gauge: Homogeneous model}.
	
	In particular the algebra of local symmetries is $\bbR \times \carr_{\gL}\left(n\right)$.
\end{Proposition}

\subsection{Poincaré operator, BMS coordinates and good-cuts}

Let $\left(\scrI \to \gS, [n^{\mu},h_{\mu\nu}], \cP \right)$ be a strong conformally Carrollian geometry. In the previous section we saw that covariantly constant dual-tractors are essentially equivalent to zeros of the Poincaré operator. We will here briefly investigate what these are in a Bondi gauge. 

Let $h_{\mu\nu} \in [h_{\mu\nu}]$ be a Bondi gauge, $u$ a BMS coordinates i.e. satisfying $du\left(n\right)=1$ and let us pick a compatible dual tetrad $\left(l = du, \gth^A\right)$. The Poincaré operator  \eqref{Null infinity: Poincaré operator} then takes the simpler form
\begin{align*}
&\cP_{00}(f)  = \covD_{0} \fd, \\
&\cP_{A0}(f)  =\covD_{A} \fd,\\
&\cP\left(f\right)_{AB}= \covD_{A}\covD_{B} \big|_{tf}\; f  +\tfrac{1}{2}C_{AB} \fd - \tfrac{1}{2} \left(\LieD_{n}C\right)_{AB} \; f,
\end{align*}
where $f$ is a function on $\scrI$. Let $f$ be a zero of $\cP$. In particular the first two equations are equivalent to
\begin{equation*}
f = k (u - \pi^*G), \qquad G\in \Co{\gS}, \quad k \in \bbR,
\end{equation*}
and the third to
\begin{equation*}
\covD_{A}\covD_{B} \big|_{tf}\; G -\tfrac{1}{2}C_{AB} +\left(u -G\right) \tfrac{1}{2} \left(\LieD_{n}C\right)_{AB} =0.
\end{equation*}
Taking an overall Lie derivative $\LieD_{n}$ one finds that $(\LieD_{n}\LieD_{n} C)_{AB} =0$ which is an integrability condition for this equation (In fact this equivalent to the vanishing of a tractor curvature coefficient and correspond to $\Psi^0_4 =0$ from the space-time point of view). Assuming this, $f = k(u-G)$ is a zero of the Poincaré operator if and only if the good-cut equation
\begin{equation*}
\covD_{A}\covD_{B} \big|_{tf}\; G - \tfrac{1}{2}C_{AB}\big|_{u=G} = 0
\end{equation*}
is satisfied.

\section*{Acknowledgement}
This project has received funding from the European Research Council (ERC) under the European Union’s Horizon 2020 research and innovation programme (grant agreement No 101002551).

\printbibliography
\end{document}